\documentclass[letterpaper,12pt]{article}
\usepackage[letterpaper, portrait, margin=1in]{geometry}
\usepackage{graphicx} 
\usepackage{makecell}
\usepackage[numbers]{natbib}
\usepackage{amsmath}
\usepackage[dvipsnames]{xcolor}
\usepackage{setspace}
\usepackage{url}
\usepackage{capt-of}
\usepackage{paralist}
\usepackage[toc,page,header]{appendix}
\usepackage{minitoc}
\usepackage[sectionbib,globalcitecopy]{bibunits}
\defaultbibliographystyle{ieeetr}
\defaultbibliography{references_all.bib} 

\begin{document}
\begin{bibunit}
\doparttoc 
\faketableofcontents 

\title{Diffusion of complex contagions is shaped by a trade-off between reach and reinforcement}
\author{Allison Wan,$^{a}$ Christoph Riedl,$^{a,b,c,*}$ David Lazer $^{a,c,d,e}$\\
\small{$^{a}$Network Science Institute, Northeastern University, Boston MA    } \\ 
\small{$^{b}$ D'Amore-McKim School of Business, Northeastern University, Boston, MA }\\
\small{$^{c}$ Khoury College of Computer Sciences, Northeastern University, Boston, MA }\\
\small{$^{d}$ Department of Political Science, Northeastern University, Boston MA     }\\
\small{$^{e}$ Institute for Quantitative Social Science, Harvard University, Cambridge MA }\\ 
\\
\small{$^{*}$To whom correspondence should be addressed; email: c.riedl@northeastern.edu }
\\
\small{Published at the Proceedings of the National Academy of Sciences}}
\date{\quad}

\begin{titlepage}
\setcounter{page}{0}
\maketitle


\begin{abstract}
\noindent How does social network structure amplify or stifle behavior diffusion? Existing theory suggests that when social reinforcement makes the adoption of behavior more likely, it should spread more---both farther and faster---on clustered networks with redundant ties. Conversely, if adoption does not benefit from social reinforcement, it should spread more on random networks which avoid such redundancies. We develop a novel model of behavior diffusion with tunable probabilistic adoption and social reinforcement parameters to systematically evaluate the conditions under which clustered networks spread behavior better than random networks. Using simulations and analytical methods, we identify precise boundaries in the parameter space where one network type outperforms the other or they perform equally. We find that, in most cases, random networks spread behavior as far or farther than clustered networks, even when social reinforcement increases adoption. Although we find that probabilistic, socially reinforced behaviors can spread farther on clustered networks in some cases, this is not the dominant pattern. Clustered networks are even less advantageous when individuals remain influential for longer after adopting, have more neighbors, or need more neighbors before social reinforcement takes effect. Under such conditions, clustering tends to help only when adoption is nearly deterministic, which is not representative of socially reinforced behaviors more generally. Clustered networks outperform random networks by a 5\% margin in only 22\% of the parameter space under its most favorable conditions. This pattern reflects a fundamental tradeoff: random ties enhance reach, while clustered ties enhance social reinforcement.

\end{abstract}

\thispagestyle{empty}
\end{titlepage}

\newpage
\section*{Introduction}

Social network structure has been shown to be consequential in the diffusion of behaviors, information, or ideas \cite{granovetter1973strength, watts1998collective, centola2007complex}. 
The question of how network structure can amplify or stifle the spread of behavior has important implications across a wide range of fields, including
health \cite{aral2017exercise,christakis2007spread},
sociology \cite{granovetter1973strength},
economics \cite{jackson2007diffusion},
organization science \cite{reagans2003network}, and 
marketing \cite{godes2009firm}.
When behaviors are sensitive to the number of influential neighbors due to strategic complementarity, social pressure, credibility, or legitimacy \cite{young2015evolution,dimaggio2012network}, they can spread better---both faster and farther---on clustered networks.
On clustered networks, neighbors of a given individual tend to be themselves densely connected, enabling repeated exposure to multiple adopting neighbors who serve as a source of social reinforcement, increasing the likelihood of adoption.
Conversely, some behaviors are not sensitive to social reinforcement. Instead, the chance of adoption is independent of how many different adopting neighbors an individual is exposed to. This is similar to the spread of biological contagions where contact with a single infected individual is sufficient to catch the disease \cite{dodds2005generalized, barrat2008dynamical}.
When behaviors are not sensitive to social reinforcement, clustering merely adds unnecessary redundancy---``wasting'' exposures on people who have already adopted---thus slowing down spread and reducing its reach \cite{watts1998collective}. 
Socially reinforced spreading processes have been called ``complex contagions'' whereas processes not sensitive to social reinforcement have been called ``simple contagions'' \cite{dimaggio2012network}.

Important parts of the prior literature on diffusion have cast this distinction about \textit{which} behavior spreads better on \textit{which} network as a sharp dichotomy: clustering facilitates the spread of socially reinforced behaviors but hinders the spread of behaviors that do not benefit from social reinforcement \cite{centola2007complex}. 
Recent work, however, suggests this dichotomy is an artifact of modeling behavior as deterministic \cite{eckles2024long, sassine2024does}. 
They find that relaxing the assumption of deterministic adoption challenges the idea that socially reinforced behavior benefits from redundant---but not random---ties.
By allowing for small amounts of probabilistic adoption (but keeping other parts deterministic), they show that socially reinforced behaviors spread faster on random networks just like non-socially reinforced behaviors.
Despite this progress, it remains unclear how general the ``faster on clustered networks/slower on random network'' dichotomy is when adoption is consistently modeled probabilistically, which is both more realistic and reflective of the empirical literature on peer influence \cite{bakshy2012role, bakshy2012social, romero2011differences, leskovec2007dynamics, lee2022complex, fink2016investigating}.
This is particularly important since some empirical \cite{centola2010spread} and theoretical studies \cite{centola2007complex} incorporating probabilistic adoption do find evidence for socially reinforced behaviors spreading farther and faster on clustered networks, apparently contradicting the results in \cite{eckles2024long, sassine2024does}. 
So \textit{why}, and \textit{under what conditions}, is clustering beneficial to the spread of behaviors when the behavior benefits from social reinforcement and is consistently modeled with probabilistic adoption?
Here, we show that the dichotomy---and the conflicting results---can be explained by a fundamental tradeoff between the benefit of redundancy that social reinforcement may harness (a \textit{reinforcement} effect) and the loss that this same redundancy causes by limiting reach to fewer individuals (a \textit{reach} effect). 
Whether clustering is harmful or helpful depends on the precise strength of both effects.
When the benefits of reinforcement outweigh the loss in reach, socially reinforced behaviors will spread better on clustered networks.  
When the reverse is true, random networks will be more beneficial.

As one way to quantify the tradeoff between reinforcement and reach, we develop a threshold model of homogeneous behavior adoption with tunable probabilistic adoption.
Socially reinforced behaviors have often been modeled with deterministic threshold models  \cite{granovetter1978threshold, schelling1969models, centola2007complex}, where exposure to a specified threshold number of influential neighbors is required for adoption. 
That is, adoption is impossible below the threshold, but guaranteed above.
When adoption is instead modeled probabilistically, surpassing the threshold of neighbors allows social reinforcement to ``kick in,'' increasing the likelihood of adoption from the below threshold base rate. Adoption becomes possible below the threshold, yet is not guaranteed above.
Our model relies on two parameters to incorporate probabilistic adoption: one tunes the base rate of adoption (below the threshold); the other tunes the socially reinforced adoption rate (above the threshold). Each parameter ranges from 0---implying no adoption above (below) the threshold---to 1---implying guaranteed deterministic adoption above (below) the threshold. This model thus efficiently captures diffusion processes in the space from simple to complex contagions.

In this paper, we use agent-based modeling and analytical techniques to analyze diffusion processes where the socially reinforced adoption rate equals or exceeds the base rate of adoption (i.e., the scenarios in which earlier literature had suggested that clustered networks should be superior at facilitating spread compared to random networks).
Additionally, we consider the effect of other relevant features of the diffusion process and network structure including the threshold of neighbors an individual must be exposed to in order to adopt at the socially reinforced above threshold rate (``social reinforcement threshold''), how long an individual remains influential after adopting (``time of influence''), and the number of neighbors each individual in the network has (``degree'').
For each parameter combination, we compare diffusion on clustered ring lattice networks \cite{watts1998collective} to that of regular random networks constructed by rewiring clustered networks \cite{gkantsidis2003markov}, while holding network size and degree constant across network types (we explore diverse variations in our robustness tests).
As outcomes, we are interested in both the reach and speed of spread.

We find there are four qualitatively distinct regions in the space defined by our two parameters: 1) clustered networks outspread random networks, 2) random networks outspread clustered networks, 3) full spread is reached on both network types but at a faster rate on random networks, or 4) there is minimal spread on both network types.
This means that there are behaviors with significant social reinforcement that spread more on random networks, as well as those that spread more on clustered networks. 
The canonical result of ``better spread on clustered networks'' in fact represents only a small portion of the parameter space. This region shrinks even more when behaviors have long times of influence, high social reinforcement thresholds, and greater network degree.
When time of influence is long, full spread dominates. Clustered networks outspread random ones only when below-threshold adoption is close to zero.

This paper makes the following contributions. 
First, our model explores a larger parameter space of simple and complex contagions, subsumes previous models as special cases, and better reflects the probabilistic nature of real world human behavior. 
Previous work that numerically simulates models with both probabilistic base rate and socially reinforced adoption did not incorporate thresholds \cite{o2015mathematical, keating2022multitype, zheng2013spreading}, and work using threshold models included probabilistic base rate adoption \cite{eckles2024long, sassine2024does} or probabilistic socially reinforced adoption \cite{centola2007complex}, but not both.\footnote{Eckles et al. \cite{eckles2024long} theoretically model diffusion processes with both probabilistic base rate adoption and social reinforcement, but only numerically simulate one case when base rate adoption is above 0 and the socially reinforced adoption rate is set to 0.5.} 
This integrative modeling framework unifies several strands of the prior literature and strengthens the consensus on the role of social reinforcement by highlighting large regions where behaviors with probabilistic adoption spread faster on random networks despite social reinforcement (aligning with the conclusions of \cite{eckles2024long, sassine2024does}) as well as regions in which socially reinforced behavior spreads better on clustered networks (supporting the conclusions of \cite{centola2007complex, centola2010spread}).
Our findings clarify that social reinforcement is a necessary but insufficient condition for clustering to be beneficial to the diffusion of behaviors.

Second, the effect of clustering is not exclusively helpful \cite{centola2007complex, centola2010spread} or harmful \cite{eckles2024long} to the diffusion of socially reinforced behaviors with probabilistic adoption as previous work suggests.\footnote{But see Sassine and Rahmandad \cite{sassine2024does}} Rather, behaviors are subject to a tradeoff between reach (from random ties) and reinforcement (from clustered ties). 
Despite its simple nature, our model offers a lens that brings together diverse studies that variously show the benefits of both random and clustered ties.

Third, we systematically explore time of influence which few studies have varied as a parameter of interest, given its challenging analytical tractability.\footnote{The closest is the ``sender inactivation'' parameter in \cite{sassine2024does} and one numerical robustness check each in \cite{centola2007complex} and \cite{eckles2024long}.} By more systematically exploring time of influence, we arrive at contradictory results compared to prior studies \cite{sassine2024does} that find time of influence only stifles spread on random networks but not on clustered networks. 
Instead, we find that shorter times of influence hinder spread on both network types.
This increases Region 1 (where clustered networks outspread random networks) and Region 2 (where random networks outspread clustered networks) at the expense of Region 3 where behaviors can spread on both network types.

Fourth, by casting the benefits of clustering as a tradeoff between reach and reinforcement, we can quantify how the regions in which clustered networks (as opposed to random networks) better spread a behavior expand or contract depending on other modeling parameters. 
 Across a wide range of parameter values, socially reinforced behaviors spread along both random and clustered ties.
Even small increases in how long individuals remain influential after adopting, how many neighbors an individual has, or how many neighbors are required to adopt at the socially reinforced rate all increase the region where full spread is reached on both network types (though at a faster rate on random networks) and decrease the region where clustered networks are advantageous.
Clustered networks only outperform random networks among behaviors with increasingly deterministic adoption, (when the base rate is close to 0 and the socially reinforced rate is close to 1) which are not representative of probabilistic complex contagions as a whole.

In summary, we find that prior work has somewhat overestimated the benefits of clustering for the spread of socially reinforced behaviors \cite{centola2007complex}.
Whereas the results reported in \cite{centola2010spread} are possible under certain conditions (and fully consistent with the complex contagions spread faster on clustered networks claim) they do not represent the dominant pattern.
For most cases, random networks spread a behavior equally or better. 
This suggests that greater diffusion on clustered networks is not a defining feature of complex contagions.

\begin{figure*}
    \centering
    \includegraphics[width =\textwidth]{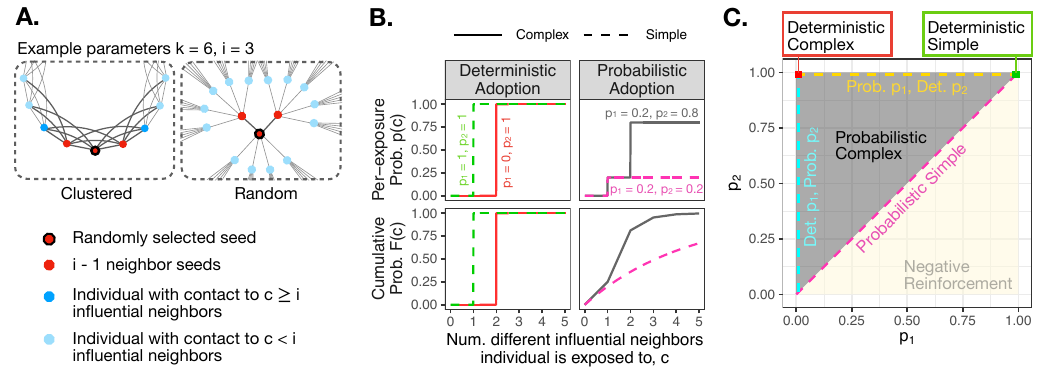}
    \caption{
    \textbf{Micro and Macro Views of Diffusion on Clustered and Random Networks.}
    \textbf{A.} The seeding structure and early time diffusion in random and clustered networks, with example parameters $k=6$ and $i =3$. Both network types are seeded by selecting one random individual and $i-1$ neighbors.
    Random networks reach more individuals ($ki-2i +2$ in the first time step), but all adopt at the lower, base rate $p_1$.
    Clustered networks reach less individuals ($k$ in the in the first time step for this example, a different set of $i-1$ neighbors could be chosen) but more individuals receive reinforcing signals, enabling adoption at the higher $p_2$ rate.
    This illuminates a tradeoff in having more or less redundant ties.
    \textbf{B.} Per-exposure and cumulative adoption probabilities for simple and complex contagions with deterministic and probabilistic adoption. 
    Example parameters $p_1 = 0.2$ and $p_2 = 0.8$ are used for behaviors with probabilistic adoption.
    \textbf{C.} The space of all possible $p_1, p_2$ values that uniquely define a behavior, or adoption trajectory ($p(c), F(c)$). Our analysis focuses on $p_1 \leq p_2$.
    }
    \label{fig:adopt_curve}
\end{figure*}

\subsection*{Establishing Micro-foundations of Social Influence: Modeling Behaviors with Probabilistic Adoption}

We introduce a model that describes the micro-level process of social influence, formalizing the differences between socially reinforced complex contagions and non-socially reinforced simple contagions. 
All individuals in the network begin having not adopted a behavior (they are ``susceptible''), except for several randomly chosen ``seed'' individuals who have already adopted and can influence their immediate neighbors to adopt (the seeds are ``infected'' individuals; Fig.~\ref{fig:adopt_curve}A).
Those who have adopted a behavior remain influential towards their neighbors for a set time length ($T$) after which they can no longer influence others (they are ``recovered''). This mirrors the Susceptible-Infective-Recovered (SIR) model from epidemiology \cite{anderson1991infectious, barrat2008dynamical}.

For each time step, all susceptible individuals are simultaneously exposed to any neighboring individuals who are currently influential. 
With every exposure to an influential neighbor, an individual may adopt the behavior with a certain ``per-exposure'' probability.
This per-exposure probability of adoption is defined by $p(c)$, where $c$ indexes the number of different influential neighbors an individual has been exposed to from the start of the simulation.
\begin{equation*}
p(c) =
    \begin{cases}
        0, & \text{if } c =0 \\
        p_1, & \text{if } 1\leq c <i\\
        p_2, & \text{if } c\geq i.
    \end{cases}
\end{equation*}
All individuals follow this adoption rule identically and there is no heterogeneity among individuals except for network position \footnote{See SI Section I for extensions with heterogeneous adoption.}.

When an individual is not in contact with any influential neighbors, they cannot adopt the behavior.
If an individual has been exposed to less than $i$ different influential neighbors, they will adopt with the base rate adoption probability of $p_1$. 
If the number of different adopting neighbors an individual is exposed to equals or exceeds $i$, the social reinforcement threshold, an individual adopts the behavior with the socially reinforced probability $p_2$.
In practice, even if an individual is exposed to multiple neighbors within one time step, the number of exposures is still counted serially.
For instance, if $i = 2$ and an unexposed individual is exposed to three influential neighbors for the first time within one time step, one neighbor ``transmits'' the behavior with the base rate probability of $p_1$ while the other two transmit the behavior with the socially reinforced probability of $p_2$.
The difference between $p_1$ and $p_2$ quantifies the amount of social reinforcement the adoption of a behavior is sensitive to, with the idea being that multiple exposures reinforce the likelihood of adoption beyond that of the base rate adoption probability $p_1$.

Setting different values of $p_1$ and $p_2$ can parameterize behaviors with different levels of base rate and socially reinforced adoption rates.
This allows us to recover well studied forms of complex and simple contagions, while at the same time allows us to examine overlooked regions of the space (Fig.~\ref{fig:adopt_curve}C). 
When $p_1 = p_2$, the threshold parameter $i$ has no effect and $p(c)$ remains constant across all additional contacts $c$.
Increasing the number of influential neighbors an individual is exposed to does not increase an individual's per-exposure probability of adoption, so the behavior is considered a simple contagion. 
The behavior is a simple contagion with deterministic adoption when $p_1 = p_2 = 1$ (or ``deterministic simple contagion''), and a simple contagion with probabilistic adoption (or ``probabilistic simple contagion'') when $ 0<p_1 = p_2 < 1$.
When $p_1 \neq p_2$, the behavior is sensitive to social reinforcement, so it is a complex contagion.
Social reinforcement can be positive when the per-exposure adoption probability increases with exposure to multiple influential neighbors, $p_1 < p_2$ (as theorized in complex contagion about costly behaviors such as attending a protest) or negative if exposure to additional influential neighbors somehow dampen each other, $p_1 > p_2$  (e.g., spreading a rumor may become less satisfying if many people already know it). 
Under both positive and negative social reinforcement, the complex contagion can have deterministic adoption ($p_1 = 0, p_2 = 1$ in the positive case; $p_1 = 1, p_2 = 0$ in the negative case) or probabilistic adoption ($0 \leq p_1 < p_2 \leq 1$ but not including $p_1 = 0, p_2 = 1$  or $p_1 = 1, p_2 = 0$). 
We focus on simple contagions and complex contagions with positive social reinforcement with deterministic or probabilistic adoption, where $p_1 \leq p_2$. 

Among complex contagions with positive social reinforcement, the social reinforcement threshold $i$ models how many different neighbors an individual must be in contact with in order to adopt at $p_2$ instead of $p_1$, ``activating'' this positive reinforcement effect.
Holding constant the total number of neighbors an individual has, or degree $k$, while increasing $i$ increases the costliness of adopting a behavior. Contact with more socially reinforcing neighbors relative to the total number of neighbors is required to adopt at the higher, socially reinforced rate $p_2$.
This is similar to other threshold models explored in the literature \cite{schelling1969models,granovetter1978threshold, valente1996social} where individuals adopt a behavior based on whether a certain threshold of neighbors adopt.
However, rather than governing deterministic adoption, surpassing $i$ only increases the likelihood of adoption from $p_1$ to $p_2$.
As we are interested in providing a minimal model that systematically varies adoption and social reinforcement, we model adoption in probabilistic terms while retaining a homogeneous social reinforcement threshold $i$ that serves as a model parameter.

The length of time an individual remains influential for after adopting, or what we refer to as ``time of influence'' $T$, models a distinction between behaviors that remain transmissible for longer periods of time as opposed to shorter periods of time. 
For instance, behaviors that remain highly visible, salient, or relevant over time (such as changing a highly visible profile picture on social media) may exhibit longer times of influence compared to behaviors where visibility quickly diminishes with time (such liking a post that quickly becomes buried under newer content).\footnote{This is similar to incorporating memory or activation time parameters into a contagion \cite{dodds2005generalized, cui2014message, sassine2024does}.}
At the extreme, such a distinction between diffusion processes with longer or shorter times of influence is analogous to the differences between the Susceptible-Infective (SI) model, where the time of influence is infinite (individuals remain influential for the entire simulation after adopting), and Susceptible-Infective-Recovered (SIR) model, where the time of influence is some finite value. 
It is well understood that SI and SIR models exhibit different diffusion patterns, suggesting that variation in time of influence may significantly affect how a behavior spreads
\cite{anderson1991infectious, barrat2008dynamical, dorogovtsev2008critical}.

Core to understanding the difference between simple and complex contagions is making a distinction between gains in diffusion from social reinforcement on the one hand, and gains from receiving repeated exposures to influential neighbors on the other.
The former, benefiting from social reinforcement, refers to an increase in the per-exposure probability of adoption of a behavior dependent on an increase in the number of different influential neighbors an individual is exposed to (adopting at $p_2$ instead of $p_1$). This is characteristic of complex contagions with positive reinforcement studied here. 
The latter, benefiting from repeated exposures to influential neighbors, refers to the extent to which the cumulative probability increases with more exposures, simply from the nature of probability (the chance of observing at least one coin toss to come up heads is higher when we flip \textit{two} coins than when flipping just \textit{one} (i.e., $p(\texttt{at least one head}) = 1 - (1-0.5)^2 = 0.75$)).
While benefiting from social reinforcement is only possible when the number of different influential neighbors exceeds the threshold $i$, benefiting from redundant exposures occurs with every exposure, regardless of whether they are from the same neighbor or different neighbors. 
Non-socially reinforced simple contagions with probabilistic adoption benefit \textit{only} from increasing exposure to influential neighbors, but complex contagions with positive social reinforcement benefit from \textit{both} redundant exposure to influential neighbors \textit{and} the socially amplified adoption probability $p_2$ (when exposures exceed the threshold $i$). 

This difference can be formalized by the cumulative probability $F(c)$ of the per-exposure probability of adoption $p(c)$, where $F_{C}(c) = P(C\leq c)$ (Fig.~\ref{fig:adopt_curve}B). 
The cumulative probability of adopting a simple contagion can be expressed as, $F(c) = 1-(1-\beta)^c$ where $p_1 = p_2= \beta$. 
When the behavior is a simple contagion with deterministic adoption, $p_1 = p_2=1$, $F(c) =1$, and the likelihood of adopting the behavior does not increase with additional exposures after the first exposure since the behavior is guaranteed to be adopted after the first exposure.
However, when $0 <\beta < 1$ and the behavior is a simple contagion with probabilistic adoption, $F(c)$ increases with additional exposures to influential neighbors, similarly to that of complex contagions, even though the behavior is not more likely to be adopted with social reinforcement.
Though in the case of complex contagions with probabilistic adoption, $F(c)$ increases at a \textit{faster} rate compared to simple contagions with probabilistic adoption with the same base rate probability $p_1$.
This is visible in the bottom right panel of Fig.~\ref{fig:adopt_curve}B: while the simple contagion with probabilistic adoption experiences an increasing cumulative adoption probability from exposure to more influential neighbors (albeit with diminishing returns), the increase for the complex contagion with probabilistic adoption is higher.

Given that both simple and complex contagions with probabilistic adoption benefit from repeated exposures through redundant ties characteristic of clustering, but can also be transmitted along non-redundant, random ties that can quickly reach disparate parts of the network, it becomes theoretically ambiguous as to whether the presence of clustering would be beneficial for spread in either case.
Probabilistic simple contagions benefit from redundant exposures, while probabilistic complex contagions with non-zero $p_1$ can benefit from reaching more unique individuals even through non-redundant ties (Fig.~ \ref{fig:adopt_curve}B). 
This stands in contrast to the clean cut deterministic case where complex contagions spread better on clustered networks because they only benefit from redundant ties, and simple contagions spread better on random networks because they only benefit from non-redundant ties.
The relative strengths of these two effects, gains from redundant ties as opposed to gains from non-redundant ties, will determine which network spreads behavior ``better''. 
By exploring this model, we will show that socially reinforced complex contagions spread farther and faster on clustered networks only in a small area of the $p_1 \leq p_2$ parameter space, whereas in the majority of the parameter space the random network performs equally or better.
We also test the effects of degree ($k$), social reinforcement threshold ($i$), and time of influence ($T$; see Methods).

\section*{Results}

\begin{figure*}[t!]
    \centering
    \includegraphics[width = \textwidth]{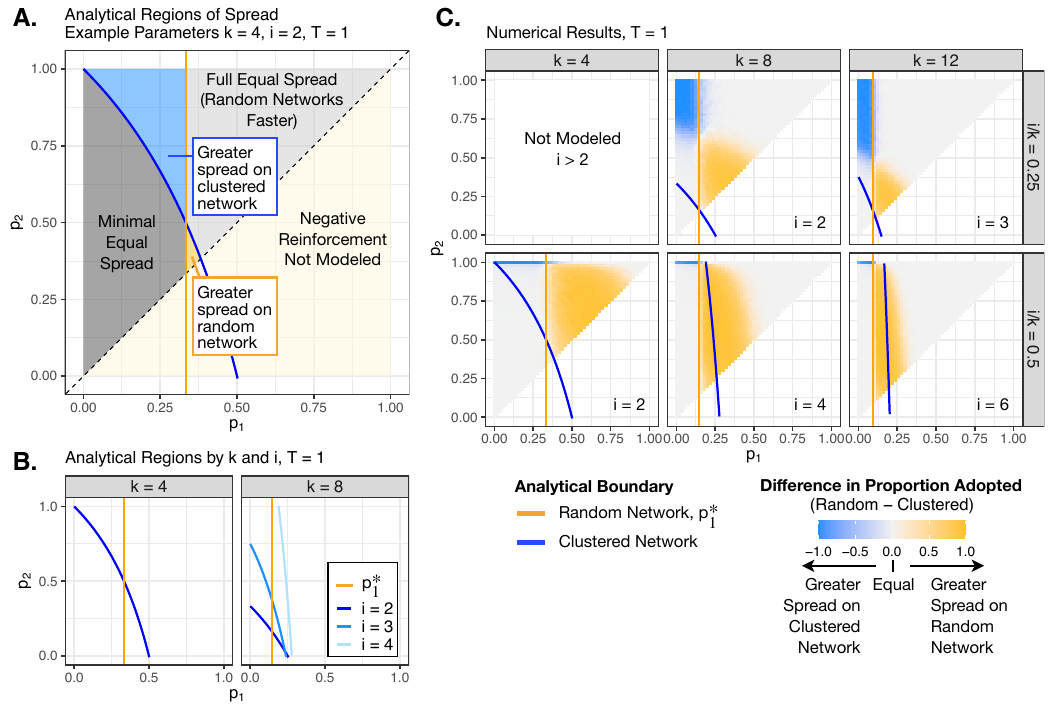}
    \caption{\textbf{Regions of spread on clustered and random networks.} 
    \textbf{A.} Analytical regions of spread on random and clustered networks with example parameters k = 4, i = 2, T = 1. The blue and orange lines denote where full spread is likely to be reached on either clustered or random networks respectively, demarcating four distinct regions in the $p_1 \leq p_2$ space: clustered networks outspread random networks (blue), clustered networks outspread random networks (orange), full equal spread on both networks with random networks faster (light gray), and minimal equal spread on both network types (dark gray).
    \textbf{B.} Analytical boundaries for different degree $k$ and social reinforcement threshold $i$.
    \textbf{C.} Analytical boundaries (colored lines) overlaid on simulation results (shaded regions, proportion of final adopters averaged over 100 trials) for different $k$ and $i$. Random networks have a greater proportion of adopters in the orange region, clustered networks have a greater proportion of adopters in the blue region.
    The analytical boundaries closely map to numerical results on random networks and underestimates values of $p_1$ and $p_2$ where full spread is reached on clustered networks.}
    \label{fig:fig2}
\end{figure*}

We observe a large amount of heterogeneity in how behaviors in the $p_1 \leq p_2$ space diffuse on clustered and random networks.
Rather than clustering being universally helpful or harmful to spread, results show four qualitatively distinct regions: 1) clustered networks outspread random networks, 2) random networks outspread clustered networks, 3) full spread is reached on both network types but at a faster rate on random networks, or 4) minimal spread on both network types.
This variation is entirely explained by the extent to which the base rate of adoption ($p_1$), socially reinforced adoption ($p_2$), network degree ($k$), time of influence ($T$), and social reinforcement threshold ($i$) modulate the tradeoff between the benefits of reach from random ties, and the benefits of social reinforcement from clustered ties.
We demonstrate this both with simulations and a simplified analytical proof (Fig.~\ref{fig:fig2}; see Methods).

On random networks, the majority of parameter combinations will either spread readily or not spread at all (SI Fig. S6). 
This creates an observable boundary in the $p_1 \leq p_2$ space where the likelihood of reaching full spread quickly transitions from very low to very high. 
This transition occurs when the base rate adoption $p_1$ is greater than a threshold value $p_1^*$, such that $p_1^* = 1 - (1 - (1/(k-1)))^{1/T}$.
If $p_1$ crosses this threshold, on average every adopting individual can influence at least one neighbor, sustaining the diffusion process. 
This reflects the logic of an epidemic threshold \cite{anderson1991infectious, dorogovtsev2008critical}.
As either network degree or time of influence increases, the chance that an adopting individual may influence their neighbor increases.
Individuals have more chances to transmit behavior, either by having more neighbors or by having more repeated transmission as the time of influence increases. 
On the macro-level, this decreases $p_1^*$, and increases the area within the $p_1 \leq p_2$ space where a behavior can attain full spread on a random network.

Spread on clustered networks is also characterized by a sharp transition from regions where spread is highly unlikely to a region where attaining full spread is highly likely.
Overcoming this transition to a region where full spread is likely on clustered networks depends on either having a high enough $p_1$ such that a behavior can spread without the benefit of social reinforcement from redundant ties, or a high enough $p_2$ such that the social reinforcement from redundant ties alone can continue the diffusion process.
When $p_1$ is too low, a behavior cannot be diffused along non-redundant ties. 
Instead, the diffusion process must be sustained along clustered ties by a high $p_2$.
As $p_1$ increases, behavior can spread more readily through both redundant and non-redundant ties, and can thus diffuse even with lower levels of $p_2$.
Once $p_1$ is high enough that a diffusion process could be sustained on clustered networks without added social reinforcement from redundant ties, full spread can be reached regardless of the value of $p_2$.
Increasing degree and time of influence both facilitate spread on clustered networks, similar to that of random networks. 
As as a result, the region where behaviors can spread on clustered networks expands to encompass lower values of $p_1$ and $p_2$.
Increasing $i$ requires contact with a greater number of adopting neighbors before social reinforcement can ``kick in.''
This makes full spread on clustered networks less likely and increases the required values of $p_1$ and $p_2$ for diffusion to occur.

\begin{figure*}
    \centering
    \includegraphics[width = \textwidth]{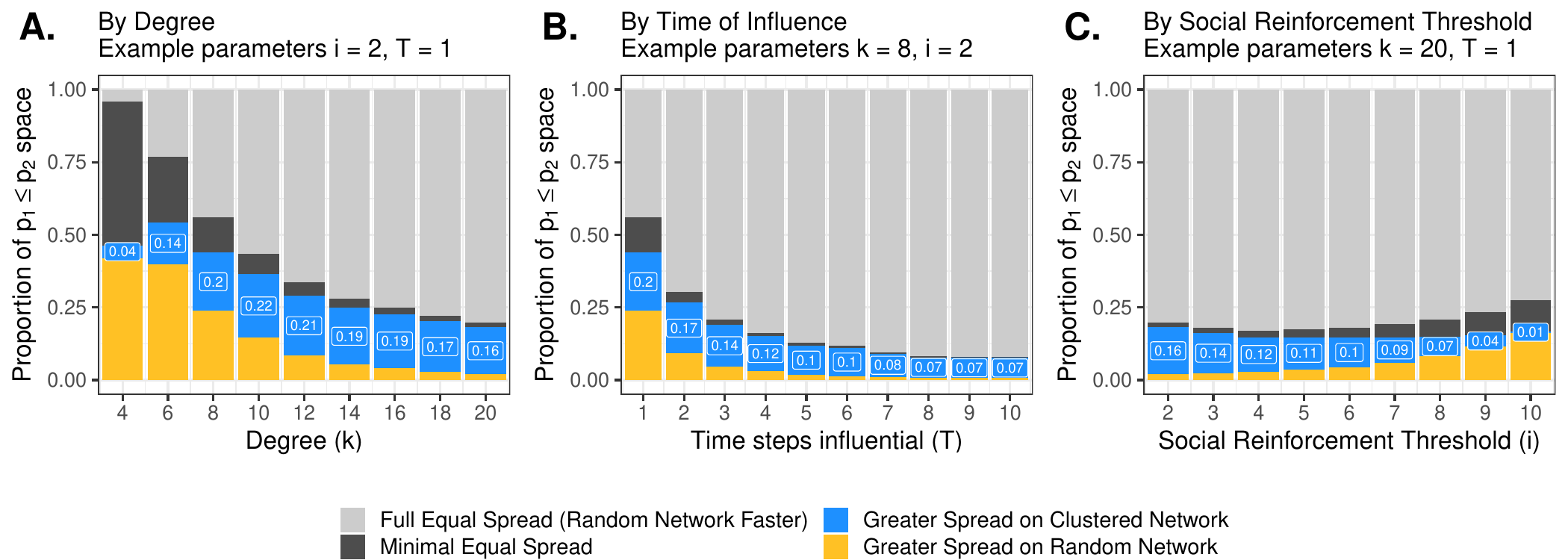}
    \caption{\textbf{Effect of Degree, Time of Influence, and Social Reinforcement Threshold on Regions of Spread in the $p_1 \leq p_2$ space.} 
    The proportion of the $p_1 \leq p_2$ where clustered networks outperform random networks (blue), random networks outperform clustered networks (orange), both network types have equal minimal spread (dark gray), or both network types have equal full spread (but faster on random networks, light gray) for varying degree (\textbf{A.}), time of influence (\textbf{B.}), and social reinforcement threshold (\textbf{C.}).  Spread, or the proportion of adopters at the end of the simulation, is averaged over 100 simulations of the same parameter combination and one network is considered to outperform another if there is at least a 5 percent difference in the proportion of adopters. Full equal spread is where the difference in the proportion of adopters between the two network types is within 5 percent and at least 60 percent of the network adopts the behavior for both network types. Minimal equal spread is where the difference in the proportion of adopters between the two network types is within 5 percent and less than 60 percent of the network adopts the behavior for either network type.
    }
    \label{fig:fig3}
\end{figure*}

Combining the forces driving spread on random and clustered networks reveals four distinct regions in the $p_1 \leq p_2$ space (Figs.~\ref{fig:fig2}A and \ref{fig:fig2}C). In Region 1, clustered networks outspread random networks. With higher $p_2$ but lower $p_2$, behaviors can spread through socially reinforcing redundant ties, but not random ties.  In Region 2, behaviors spread on random networks, but not clustered networks.  Behaviors have a high enough $p_1$ to overcome the $p_1^*$ threshold, spreading readily on random networks. However, neither $p_1$ nor $p_2$ is high enough for behaviors to spread on clustered networks. In Region 3, $p_1$ and $p_2$ are both high enough that behaviors can fully diffuse on either random or clustered networks. That being said, the ability of random ties to reach many individuals through short paths (when few individuals serve as intermediaries between any two individuals in the network), means that behaviors reach full spread at a faster rate on random networks compared to clustered networks in this region. Finally in Region 4, both $p_1$ and $p_2$ are low, so there is minimal spread on both network types.  Neither the redundant ties of clustered networks nor the ability for random networks to reach many unique individuals can be used advantageously. As random networks will always diffuse a behavior faster in Regions 2 and 3, focusing on each of these regions can describe both the speed and reach of spread. Our analytical model maps closely to the boundary where full spread is reached on random networks, and serves as an informative estimate for where full spread on clustered networks is reached (SI Fig. S1). Moreover, this analytical model accurately reflects how variables including network degree, time of influence, and the social reinforcement threshold affect spread.

Altogether, Region 1 where clustered networks outperform random networks constitutes a minority of the $p_1 \leq p_2$ space that approaches a deterministic setting(low $p_1$, high $p_2$).
Region 1 is maximized and encompasses more probabilistic regions of the space when time of influence is short, degree is low, and the social reinforcement threshold is low. 
This most likely reflects the parameter settings characteristic of the empirical behaviors studied in \cite{centola2010spread} which found faster spread on clustered networks (for a more detailed comparison see SI Section L).
Even under these ideal conditions, across all the parameter combinations we examine, clustered networks spread a behavior to five percent more of the total network compared to random networks for at most only 22 percent of the $p_1 \leq p_2$ space (for the parameter combination $k=10$, $T = 1, i = 2$; Fig.~\ref{fig:fig3}A).\footnote{We select a difference in spread of five percent of the total individuals in the network as a threshold for substantive significance. For instance, on networks with $k=4$ and 1000 individuals this would be a difference of 50 individuals. We additionally test for statistical differences using a non-parametric Kolmogorov–Smirnov(KS) test. Although a KS test slightly increases the area where clustered networks outspread random networks, this region still constitutes a minority of the total space and shrinks with greater degree, time of influence, and social reinforcement threshold. See SI Section C.}
The minimal amount of social reinforcement (ratio of $p_2$ to $p_1$, $p_2/p_1$) we observe where clustered networks spread a behavior to at least five percent more of the network compared to random networks occurs when $p_2$ is nearly two times that of $p_1$ ($p_2/p_1 =1.90$; SI Fig. S4).
This is important insofar that among empirical studies of peer influence, it is rarely the case that the ratio of socially reinforced to non-socially reinforced adoption rates ever exceed two (see \cite{eckles2024long, lee2022complex} for reviews).

With the exception of networks with low degree where the substantive difference in spread between the two networks types is smaller ($k = {4,6,8}$), greater degree, time of influence, and social reinforcement all shrink Region 1 towards more deterministic areas of the $p_1 \leq p_2$ space (Fig.~\ref{fig:fig3}).
First, increasing the social reinforcement threshold raises the number of different neighbors an individual must be exposed to in order to adopt at the $p_2$ rate, requiring a $p_2$ closer to 1 to sustain diffusion on clustered networks.
This shrinks Region 1, making it more deterministic. 
Second, interventions that increase spreading on both clustered and random networks (such as greater degree or longer time of influence) decrease $p_1^*$ on random networks and enable spread on clustered networks for lower $p_1$ and $p_2$ values.
As a result, Region 3 where both network types reach full spread, though at a faster rate on random networks, expands at the expense of Regions 1 (clustered networks spread behaviors better) and 2 (random networks spread behaviors better; Fig.~\ref{fig:fig2}, SI Fig. S6).
Region 1 narrows to smaller, more deterministic values of $p_1$, but also extends to lower values of $p_2$.

Since the shrinkage from smaller $p_1$ outpaces the gains with $p_2$, Region 1 shrinks with greater degree and time of influence.
By extending time of influence so that adopters remain influential for the entire simulation as modeled in \cite{centola2007complex, sassine2024does, eckles2024long}, aside from where $p_1$ approaches 0 and $p_2 > 0$ where clustered networks perform best, Region 3 dominates (SI Fig. S6E). 
This means for deterministic $p_1 = 0$, clustered networks outperform random networks for any amount of (deterministic or probabilistic) $p_2 > 0$, explaining the results of \cite{centola2007complex} who use the point $p_1 =0, p_2 = 0.5$ as a robustness check for probabilistic adoption (See SI Section K).
However, virtually any increase in $p_1$ from 0 would push a behavior into Region 3 where full spread is attained equally on both clustered and random networks, but at a much faster rate on random networks consistent with the core results of \cite{eckles2024long}.

Finally, our conclusions about the effects of time of influence diverge somewhat from those of \cite{sassine2024does} who incorporate a similar parameter. 
Whereas they find that shorter times of influence advantage clustered networks (over random networks), they conclude this is because short times of influence stifle spread on random networks while clustered networks remain unaffected. 
Instead, we find that short times of influence limit spread on both network types, but for different parts of the $p_1 \leq p_2$ space. This insight derives from modeling all combinations of $p_1 \leq p_2$ instead of merely $0 \leq p_1 \leq 1, p_2 = 1$ as \cite{sassine2024does} do.
For instance, if we consider networks where $k =4$, there is no difference in spread on clustered networks from $T=1$ (SI Fig. S6A) to a longer time of influence ($T=5$) when $p_2 = 1$ (SI Fig. S6C).
However, there is a dramatic change in spread from $T=1$ to $T=5$, when $p_2 < 1$.

\section*{Discussion}
Over the past 15 years, the theory of complex contagion has emerged as one of the most influential and paradigm-shifting ideas in the study of social influence \cite{dimaggio2012network}. In contrast to the long-standing belief in the structural advantages of random ties for spreading information or behaviors quickly and widely \cite{granovetter1973strength, watts1998collective}, the complex contagion argument introduced a fundamentally different perspective. It posited that the adoption of costly behaviors---those requiring socially reinforcing signals from multiple neighbors---is more effectively supported by the redundant ties found in clustered networks \cite{centola2007complex}. Whereas the theory was originally developed in a deterministic context, it assumed that such findings would generalize to a probabilistic setting which more accurately captures the capriciousness of human behavior. 

We show that the apparent dichotomy about which behavior spreads faster on which network can be explained by a fundamental tradeoff between a reinforcement effect and a reach effect. Whether clustering is helpful or harmful to the spread of socially reinforced behavior depends on the precise strength of both effects. 
As a result, both results reported in the prior literature are possible: in one region of the parameter space between base rate and social reinforcement, behavior spreads better on clustered networks; in another region, behavior spreads better on random networks.  
The region in the parameter space where reinforcement outweighs reach is small, resembles the deterministic setting (where $p_1$ is close to $0$ and $p_2$ close to $1$), and unlikely to occur in real world diffusion processes. 
In the majority of cases, both networks spread a behavior equally far (though at a faster rate on random networks). 
We provide analytical bounds that divide the parameter space into regions in which each case occurs. 
Our numerical analyses additionally explore how the different regions expand or shrink when behaviors have long times of influence, high social reinforcement thresholds, and network degree is high. The general pattern is this: once we allow for a shift from social reinforcement being \textit{necessary} before adoption can occur to merely being \textit{amplified}, diffusion processes cease to depend only on clustered ties and begin to benefit from random ties. 

Our work builds on prior work, notably \cite{sassine2024does} and \cite{eckles2024long} which made important contributions regarding stochastic diffusion processes. We add to the cumulative picture of what we know about complex and simple contagions. We quantify for which kinds of diffusion processes (in terms of base adoption rate and strength of social reinforcement) clustered as opposed random networks spread a behavior better, how the different regions expand or shrink as other parameters change (degree, time of influence, and threshold), and provide important robustness tests (heterogeneity in adoption, one- and two-dimensional clustered networks, alternative seeding strategies, and intermediate levels of rewiring). We do so in a more systematic way, exploring both probabilistic above and below threshold adoption.

Regarding other models that have explored probabilistic spreading (e.g., \cite{centola2007complex}) we find that when adopters remain influential for the entire simulation, spread on clustered networks is sensitive to how close to deterministic the base rate adoption is rather than how deterministic above threshold adoption is. Some past work has thus overlooked the important effects of non-deterministic base rate adoption which results in faster spreading on random networks and not clustered networks.\footnote{\cite{eckles2024long, sassine2024does} are notable exceptions.} Regarding \cite{centola2010spread}, our synthetic replication reveals that there is indeed a region in the $p_1 \leq p_2$ space where socially reinforced behavior with probabilistic adoption spread only on clustered networks. However, for many different levels of $p_1$ and $p_2$ there are also results showing spread exclusively on random networks, spread on both networks types, or no spread at all. The study's choice of two dimensional networks, networks with low degree, and the transmission of a behavior with a presumably low time of influence all serve to maximize the region where clustered networks are beneficial, increasing the likelihood of finding the result he did. The advantage of clustered networks is strongest when social reinforcement is \textit{necessary} for adoption to occur, yet is substantially weaker when it merely amplifies adoption. In that case, random ties increase spread substantially.

At best, clustered networks only outperform random networks by at least five percent when social reinforcement is at least two times that of the baseline probability of adoption.
Based on this, among empirical studies of peer influence or social contagion, it is rarely the case that the socially reinforced adoption rate is even proportionally high enough for clustered networks to outperform random networks (see \cite{eckles2024long, lee2022complex} for reviews).
For most realistic, probabilistic human behavior, a random network will generally spread a behavior just as well, if not better, than a clustered network. 

In summary, we develop a simple two-parameter model for simple and complex contagions that subsumes many prior models as special cases. We provide analytic bounds that demonstrate large amounts of heterogeneity in how well random and clustered networks diffuse socially reinforced behaviors. Our work bridges theory developed for deterministic and probabilistic threshold diffusion processes, painting a richer, cumulative picture of possible outcomes, resulting from a tradeoff of reach (benefit of random ties) and reinforcement (benefit of clustered ties). 

Our work is not without limitations. Whereas we explore a variety of different network structures, including one- and two-dimensional lattices and those with both clustering and short paths, future work could explore networks with heterogeneous degree  \cite{barash2012critical, centola2008failure}, or on empirical social networks. Our work is mostly focused on homogeneous individual behavior (although we have important robustness tests using heterogeneous adoption), and we do not examine potential correlations between clustering and other features that may affect diffusion including the relational strength of ties between individuals \cite{krackhardt2003strength,onnela2007structure}, heterogeneity in the number of neighbors an individual has \cite{barash2012critical,centola2008failure,guilbeault2021topological}, or homophily of individual traits \cite{centola2011experimental, centola2013homophily, melamed2020homophily, aral2009distinguishing}.
All would be fruitful directions for future work.

\section*{Methods}
\subsection*{Network Structure}
 In all networks, individuals have the same degree $k$. The number of individuals in the network is scaled to $n = 250k$ to preserve network density and isolate the effect of degree. Networks with $k=4$ have 1000 individuals, those with $k=8$ have 2000 individuals, and so on.
Clustered networks are Watts-Strogatz style ring lattices \cite{watts1998collective} where each node is connected to their $k$ nearest neighbors.
Random networks are generated by taking the original ring-lattice and performing a series of degree- preserving edge swaps according to the procedure outlined in \cite{gkantsidis2003markov} and implemented in networkX.
Specifically, two random edges, say $(u,v)$ and $(x,y)$, are selected and swapped to become $(u,x)$ and $(v,y)$
where the number of swaps is equal to the number of edges. 
If the edge swap disconnects the network, or if the new edge created with the swap already exists, the swap is rejected. 
For computational efficiency, the algorithm only checks for connectedness after every three swaps.
This procedure uniformly samples the space of connected regular networks with a specified degree sequence \cite{gkantsidis2003markov}.
The resulting networks are regular random networks where all individuals have the same number of neighbors. 
This is similar to procedures used in related work \cite{centola2007complex, centola2010spread, sassine2024does, zheng2013spreading}.

\subsection*{Adoption Rule}

Individuals adopt the diffusing behavior with either $0, p_1,$ or $p_2$ probability based on whether the number of influential neighbors they are in contact with, $c$, has overcome the social reinforcement threshold $i$. 
\begin{equation*}
p(c) =
    \begin{cases}
        0, & \text{if } c < 1\\
        p_1, & \text{if } 1\leq c <i\\
        p_2, & \text{if } c\geq i
    \end{cases}
\end{equation*}
The number of contacts is cumulatively counted from the beginning of the simulation, rather than at the beginning of each time step. In other words, individuals only need to be exposed to $c\geq i$ different influential neighbors over the course of the entire simulation instead of simultaneously at one time step to overcome $i$. This modeling choice was made because it may more realistically represent adoption behavior, such as that in \cite{centola2010spread}, especially when time of influence is short.
Additionally, multiple exposures in one time step are still counted serially. If $i =2$, and an individual is exposed to three influential neighbors for the first time in one time step, one neighbor ``transmits" the likelihood of adoption with a rate of $p_1$ while the other two, and all subsequent exposures at later time steps with a rate of $p_2$.

\subsection*{Seeding and Diffusion}
At the start of the simulation, we seed the network with $i$ individuals who have already adopted and are influential at the first time step. 
This is done to ensure that within the first time step there is at least the possibility that a neighboring potential adopter will overcome the social reinforcement threshold and adopt with $p_2$.
For both clustered and random networks, one randomly chosen individual and a randomly selected set of $i-1$ of its $k$ neighbors are chosen as the original seeds.
The seed individuals remain influential for $T$ time steps.
We consider the effects of alternative seeding strategies in the SI Section H.

Once an individual adopts a behavior, they remain influential for $T$ time steps. After this, they are no longer influential, nor can they un-adopt and then re-adopt. This is similar to being ``recovered'' or ``removed'' class in the Susceptible-Infective-Recovered (SIR) model \cite{anderson1991infectious}. Individuals can also remain influential for the entire duration of the simulation after adopting (effectively having infinite $T$), similar to the Susceptible-Infective (SI) model \cite{anderson1991infectious}.
Simulations run until a steady state is reached either when there are no more influential individuals, or no more susceptible individuals. 

\subsection*{Simulation Trial Structure}
For each parameter combination ($p_1, p_2, k, T, i$, and additional parameters covered in the Supplementary Information) the simulation is run on both the clustered and rewired regular random network 100 times. 
A different random network and starting seed set is used for each of the 100 trials.
The clustered network remains unchanged from trial to trial because it is deterministically constructed and seeded.

The outcomes of interest include the reach of spread, measured as the proportion of individuals having adopted the behavior at the end of the simulation, and speed of spread, measured as the number of times steps required for 60 percent of the network to adopt a behavior (See SI Section E for a more detailed discussion). 
Both the amount of spread and speed outcomes are averaged over the 100 trials on each parameter combination.

\subsection*{Analytical Proof}

We develop an analytical framework to determine the values of base rate $p_1$ and socially reinforced above threshold adoption rates $p_2$ for which full spread is likely to be reached as a function of network degree ($k$), social reinforcement threshold ($i$), and time of influence ($T$).
To do so, we simplify the case to deriving the expected number of individuals, or ``initial adopters" $\langle I\rangle$, that will adopt from contact with the initial set of seed individuals in the first $T$ time steps on either regular random or clustered ring lattice networks.
Values of $p_1$ and $p_2$ where $\langle I\rangle$ equals or exceeds the number of initial seeds, which for our purposes is set equal to the social reinforcement threshold $i$, have a high chance of reaching full spread.
Knowing exactly where $\langle I(p_1, p_2, k, i) = i\rangle$ creates a boundary between regions where full spread is either likely (when $\langle I\rangle \geq i$) or unlikely to be reached (when $\langle I\rangle < i$). 

Since we are deriving expected values of initial adopters, there is still the possibility that full spread is not attained when $\langle I\rangle \geq i$  or when full spread is attained when $\langle I\rangle < i$.
However, this probability is low considering that for most parameter combinations the likelihood of a diffusion process reaching full spread is either really high or low (SI Fig. S6). 
Qualitatively, this creates a visible boundary in the $p_1 \leq p_2$ space where full spread is highly unlikely to where full spread is highly likely.
On random networks, $\langle I\rangle =i$ maps closely to the boundary between minimal and full spread observed in simulations. 
For clustered networks, $\langle I\rangle =i$ underestimates the boundary observed in simulations.
Despite this imprecision, knowing where $\langle I\rangle =i$ in the $p_1 \leq p_2$ space allows us to accurately model how $p_1, p_2, k, T$ and $i$ affect spread on random and clustered networks.


Finding $\langle I\rangle$ requires tracing the number of individuals who are exposed to at least one seed individual.
Moreover, because adopting at either below or above threshold rates depends on the number of contacts to different seeds a particular individual has, we must know the number of individuals having contact with exactly $a$ seed individuals where at the least $a=0$ and at most $a$ is equal to the number of seeds $i$. 
We define $j_a$ as the number of individuals having contact with exactly $a$ seeds. 

When the number of seeds an individual is exposed to is below the social reinforcement threshold ($a < i$), the independent probability of adopting at each contact is the base rate, $p_1$. 
If seeds are only influential for one time step ($T=1$), the cumulative probability of adoption with each successive contact where $a<i$ is $1-(1-p_1)^a$. 
If seeds are influential for multiple time steps ($T>1$), this can be rewritten as $1-(1-p_1)^{Ta}$.
Seeds may continuously influence their neighbors for each time step they are influential for, but only at the base rate $p_1$.

When the number of seeds an individual is exposed to is equal to that of the social reinforcement threshold ($a = i$), when $T =1$, the cumulative probability is $1-(1-p_1)^{i-1}(1-p_2)$. 
Generalizing to cases where $T>1$, this expression becomes $1-(1-p_1)^{i-1}(1-p_2)^{Ti-(i-1)}$.
For the first $i-1$ exposures to a seed, an individual adopts with rate $p_1$. All subsequent exposures ($Ti-(i-1)$) are adopted at the above threshold rate $p_2$.
Together, the cumulative probability of adoption, $F_a$ can be expressed as:

\begin{equation*}
F_a =
    \begin{cases}
        1-(1-p_1)^{Ta}, & \text{if } a < i\\
        1-(1-p_1)^{i-1}(1-p_2)^{Ti - (i-1)} & \text{if } a = i.
    \end{cases}
\end{equation*}
    

The sum of the product between the cumulative probability ($F_a$) and the number of individuals in contact with exactly $a$ contacts ($j_a$) for each value of $a$ yields the total expected number of initial adopters, $\langle I\rangle$.

\begin{align*}
    \langle I \rangle &= \sum_{a=1}^{i}F_a j_a\\
    &= \sum_{a=1}^{i-1}(1-(1-p_1)^{Ta})j_{a} + \\&\qquad\sum_{a=i}^{i}(1-(1-p_1)^{i-1}(1-p_2)^{Ti -(i-1)})j_{a}\\
    &= \biggl[ \sum_{a=1}^{i-1}(1-(1-p_1)^{Ta})j_{a} \biggr] + \\&\qquad(1-(1-p_1)^{i-1}(1-p_2)^{Ti -i+1})j_{a=i}
\end{align*}
Determining $j_a$, or the number of individuals in contact with exactly $a$ seeds depends on the specific network structure.

\subsubsection*{Random Networks}

For analytical simplicity, we will assume random networks are seeded randomly rather than by choosing one individual and $i-1$ neighbors as in the main results. 
Robustness checks show there is minimal difference in spread between these seeding strategies (SI Section H).
We additionally assume that networks are sparse such that the degree is small relative to the number of individuals in the network, $k<<n$.
Under these assumptions, we show that as $n \to \infty$, $j_{a=1} \to ik$ and $j_{a>1} \to 0$.
That is, there will be $ik$ individuals that are connected to exactly one seed ($k$ individuals for each seed) and no individuals connected to more than one seed. 

There are two cases where there could be less than $ik$ individuals that are connected to exactly one seed node. 
Either multiple seed nodes are adjacent to each other (case 1) or seed nodes share the same neighbor (case 2). We will show that the likelihood of either case occurring approaches zero as $n \to \infty$. 

For case 1, as $n \to \infty$, each seed has $k$ edges to possibly connect to one of the other $i-1$ seeds. 
Since any given network is uniformly sampled from the space of all possible connected regular random networks, any individual has an equally likely chance to be connected to any other individual. 
This means any seed has $k$ independent, equally likely chances of connecting to any of the other $n-1$ individuals in the network.
If there are $i-1$ other seeds, there is a $(i-1)/(n-1)$ chance that one of the $k$ edges of the target seed will connect to another seed.
For large $n$ and small numbers of seeds $i$, this can be modeled as a binomial draw where there are $k$ chances for a seed node to connect to one of the other $i-1$ seeds with probability $(i-1)/(n-1)$, $Bin(k, \frac{i-1}{n-1})$. 
The expected number of neighbors of a seed that are themselves seeds is $k(i-1)/(n-1)$, which approaches 0 as $n \to \infty$.

For case 2, as $n \to \infty$, the probability that a non-seed neighbor of a seed is also connected to at least one other seed also approaches 0. 
Any neighbor of a seed has $k-1$ edges that could connect to at least one of the $i-1$ seeds out of $n-2$ nodes (total number of nodes without the original seed and seed neighbor).
By the same logic as case 1, this can be modeled as a binomial draw where a node has $k-1$ chances of selecting one of the other $i-1$ seeds with probability $(i-1)/(n-2)$,$Bin(k-1, \frac{i-1}{n-2})$.
The expected number of seeds this neighbor of the original seed is connected to is $(k-1)(i-1)/(n-2)$, which also tends towards 0 as $n \to \infty$.

In sparse networks, any seed is unlikely to be neighbors with a seed or to share neighbors with another seed. 
This means each of the $i$ seeds are connected to $k$ unique neighbors, such that there are $ik$ total individuals connected to exactly one seed and no individuals connected to more than one seed.
The expected number of initial adopters will be that where $j_{a=1} = ik$ and $j_{a>1} = 0$. Since $i>1$, $j_{a =i} =0$.
\begin{align*}
    \langle I_R \rangle &= \sum_{a=1}^{i}F_a j_a\\
    &= \biggl[ \sum_{a=1}^{i-1}(1-(1-p_1)^{Ta})j_a\biggr] + 
    \\&\qquad (1-(1-p_1)^{i-1}(1-p_2)^{Ti -(i-1)})j_{a=i} \\
    &= (1-(1-p_1)^{T(1)})(j_{a=1}+j_{a=2}...j_{a=i-1}) +  \\&\qquad(1-(1-p_1)^{a-i}(1-p_2)^{Ti -(i-1)})j_{a=i} \\
    &= (1-(1-p_1)^{T(1)})(ik+0...0) + 
    \\&\qquad (1-(1-p_1)^{a-i}(1-p_2)^{Ti -(i-1)})(0) \\
    &=  (1-(1-p_1)^{T})ik\\
\end{align*}
To continue the diffusion process, in expectation there must be one initial adopter for each seed. 
This logic is similar to the epidemic threshold
\cite{anderson1991infectious}. 
Within the first $T$ time steps the seeds are influential for, the $i$ seeds must transmit the behavior to at least $i$ other individuals to continue the diffusion process, so 
\begin{align*}
\langle I_R \rangle =(1-(1-p_1)^{T})ik &\geq i\\ 
1-(1-p_1)^{T} &\geq \frac{i}{ik}\\ 
1-(1-p_1)^{T} &\geq \frac{1}{k}
\end{align*}
Beyond the initial seed nodes, an influential individual must influence at least one neighbor out of their $k-1$ neighbors who have not yet adopted (assuming one neighbor has already adopted in order to pass on the behavior to the currently influential individual). 
This means $1-(1-p_1)^{T} \geq 1/(k-1)$, from which we calculate $p_1^*$, or the value that the base rate $p_1$ must be equal or exceed in order for diffusion to be sustained on random networks.
\begin{align*}
p_1 &\geq 1- \bigg(1 - \frac{1}{(k - 1)}\bigg)^{1/T}\\
p_1^* &= 1- \bigg(1 - \frac{1}{(k - 1)}\bigg)^{1/T}\\
\end{align*}
The equation for $p_1^{*}$ shows that spread on random networks only depends on  $p_1$, $k$, and $T$. 
The larger either $k$ or $T$, the smaller $p_1^*$ can be, increasing the area within $p_1 \leq p_2$ where full spread is likely attained on random networks. 

\subsubsection*{Clustered Networks}

In clustered networks, as a simplifying assumption, we pick $i$ adjacent individuals as the initial seeds instead of one randomly selected individual and $i-1$ randomly selected neighbors.
We observe minimal differences between these two seeding strategies (SI Fig. S8).
Specific to selecting adjacent seeds, 
\begin{equation*}
j_{a} = 
    \begin{cases}
        2, & \text{if } a < i\\
        2(\frac{k}{2}-i+1), & \text{if } a = i
    \end{cases}. 
\end{equation*}
By the same logic as the random network, the overall number of initial adopters can be expressed as 
\begin{align*}
    \langle I_L \rangle &= 2\sum_{a=1}^{i-1}(1-(1-p_1)^{Ta}) + \\&\qquad(1-(1-p_1)^{i-1}(1-p_2)^{Ti-i+1})(k-2i+2)
\end{align*}
Setting $\langle I_L \rangle= i$
and solving for $p_2$, 
\begin{align*}
    p_2 =\frac{2\sum_{a=1}^{i-1}(1-(1-p_1)^{Ta})- 3i + k+2}{(k-2i+2)(1-p_1)^{i-1}}
\end{align*}
While the $\langle I_L \rangle= i$ boundary accurately models the regions of the $p_1 \leq p_2$ space for which the average number of initial adopters equals or exceeds the number of seeds ($\langle I_L \rangle \geq i$), it underestimates the values of $p_1$ and $p_2$ for which full spread is likely reached on clustered networks (SI Fig. S1).
Despite this discrepancy, the effect of $k$, $i$, and $T$ on the $\langle I_L \rangle \geq i$ boundary, mirrors the effect of these parameters on in the simulated results.
That is, as $k$ and $T$ increase, diffusion on clustered networks becomes more likely at increasingly lower values of $p_1$ and $p_2$.
As $i$ increases, diffusion becomes less likely as contact to a greater number of influential neighbors are required for an individual to adopt with the higher $p_2$ rate.

\section*{Code Availability}
 Code for simulations have been deposited in OSF (\url{https://osf.io/8jfcy/?view_only=dc9a6a759be941caa618621e 297a483b)}.
 
\section*{Acknowledgments}
We wish to thank Jessica Davis, Guillaume St-Onge, Cory Glover, and Alexi Quintana Math\'{e} for helpful comments.
We also thank two other anonymous referees, and editors at PNAS for their helpful feedback in revising the manuscript.
\clearpage
\singlespacing
\putbib
\end{bibunit}

\clearpage
\appendix

\setcounter{figure}{0}    
\setcounter{table}{0}
\setcounter{page}{0}

\renewcommand\thefigure{S\arabic{figure}} 
\renewcommand\thetable{S\arabic{table}} 
\renewcommand\thepage{S\arabic{page}}

\noindent \LARGE{\textbf{Supplementary Information for:}\\ ``Diffusion of complex contagions is shaped by a trade-off between reach and reinforcement''} 

\thispagestyle{empty} 

\part{} 
\parttoc 

\clearpage
\normalsize

\section{Review of Related Work}\label{si_lit_rev}

Despite the proliferation of complex contagion as a term to describe behaviors that benefit from social reinforcement \cite{karsai2014complex, ugander2012structural, traag2016complex, barash2012salience,romero2011differences, fink2016complex,lee2022complex,steinert2017spontaneous}, a small number of studies critically assess whether such behaviors diffuse more on clustered networks compared to random networks. 
The study by Centola \cite{centola2010spread} is one of the few, if not the only, controlled field experiments manipulating the effect of clustering and finds that complex contagions spread farther and faster on clustered networks compared to random networks. Other work consists mainly of simulation or analytical studies that vary in outcome measures and conclusions, depending on how the model and parameter space are specified \cite{eckles2024long, sassine2024does, keating2022multitype, lu2011small, cui2014message, zheng2013spreading, de2009role}.
We will focus our review on the subset of modeling work (simulations or analytical) that examines the effect of probabilistic adoption on how socially reinforced behaviors spread on random and clustered networks.

Among the models closest to the one presented here, there are two main ways in which probabilistic adoption has been incorporated. 
First, adoption can be modeled with a threshold model where an individual must be exposed to a threshold number of different neighbors to adopt at the above threshold, socially reinforced adoption rate rather than the below threshold, base rate.
Some models only feature probabilistic base rate adoption \cite{sassine2024does,eckles2024long}, 
with deterministic above threshold adoption.
Others only feature probabilistic above threshold, socially reinforced adoption \cite{centola2007complex} but deterministic base rate adoption.
The second common case are models that include both probabilistic base rate and socially reinforced adoption\cite{keating2022multitype, zheng2013spreading}, but without a defined threshold. 
Instead, the adoption trajectory is a smooth function that depends on base rate adoption, social reinforcement, and increases with the number of different neighbors an individual is exposed to.
We are unaware of any work that sweeps the space of threshold models with both tunable, probabilistic base rate and socially reinforced adoption rather than selecting specific points.\footnote{Both Eckles et al. \cite{eckles2024long} and De Kerchove et al. \cite{de2009role} use a threshold model and explore a few points where both base rate adoption and socially reinforced adoption are probabilistic.} 
Most models find cases where random networks diffuse a behavior better than clustered networks even when the behavior benefits from social reinforcement \cite{eckles2024long, sassine2024does, keating2022multitype, o2015mathematical, zheng2013spreading, de2009role}. 
However, there are disagreeing conclusions around how ``typical'' such cases are to socially reinforced behaviors, often because researchers are selecting specific values for parameters rather than systematically sweeping the space \cite{centola2007complex, sassine2024does}.

Two other modeling decisions that may lead to differing outcomes are varying the threshold number of neighbors required to adopt at the socially reinforced rate, or ``social reinforcement threshold,'' as well as the length of time an individual is influential for, which we call ``time of influence''. 
Varying the social reinforcement threshold reflects a theoretical distinction in how costly a behavior is.
Behaviors that require exposure to a greater number of influential neighbors relative to the number of total neighbors an individual has are costlier.
Besides Centola and Macy \cite{centola2007complex}, existing work has largely overlooked the effect of varying social reinforcement thresholds \footnote{But also see \cite{eckles2024long} on the effect of different social reinforcement thresholds.}. Some models feature fixed adoption thresholds\cite{sassine2024does}. Others do not have threshold dynamics at all, where instead the probability of adoption increases smoothly with additional exposure to influential neighbors as a function of the base rate adoption and a social reinforcement parameter \cite{keating2022multitype, zheng2013spreading}. 

Variation in time of influence often depends on the affordances of communication between individuals in a network.
For instance, behaviors that remain visible or salient for long periods of time (such as changing an account profile picture on social media) may have a long time of influence and exhibit dynamics closer to that of the Suceptible-Infective (SI) model from epidemiology where individuals remain infective for the entire course of the simulation \cite{anderson1991infectious, dorogovtsev2008critical}. 
Other behaviors (such as liking a social media post that becomes buried under new content within a few days) may have a shorter time of influence and exhibit dynamics more similar to a Suceptible-Infective-Recovered (SIR) model where individuals are only infective for a limited period of time.
This distinction matters insofar as conclusions drawn from simulations in which individuals always remain influential (SI model) have been applied to behaviors that may have shorter times of influence \cite{centola2010spread}.
Moreover, epidemiological work shows that there are large differences between the diffusion of SI and SIR models \cite{dorogovtsev2008critical}.
Among simulation studies most closely related to our work (comparing diffusion on clustered and random networks) the focus is primarily on either SI \cite{eckles2024long, centola2007complex} or SIR \cite{de2009role, zheng2013spreading, o2015mathematical,keating2022multitype} style models, but not both.\footnote{The closest is the ``sender inactivation'' in Sassine and Rahmandad \cite{sassine2024does}, and one robustness check in Eckles et al. \cite{eckles2024long}} 


For our paper, we extend existing work in several ways (Table \ref{tab:summary1} \& \ref{tab:summary2}). First, to avoid potential selection bias in picking a few probabilistic values for either base rate or socially reinforced adoption, we do a sweep of all possible combinations of these two parameters, where socially reinforced adoption either equals or exceeds the base rate. Second, we incorporate a variable social reinforcement threshold parameter to test how the costliness of a behavior affects its ability to spread on random and clustered networks. Finally, we vary the time for which an individual is influential. 
We also include additional robustness checks in the remainder of this supplement assessing the effects of heterogeneous adoption probabilities, intermediate rewiring, different seeding strategies, and higher dimensional clustered networks. 


\begin{table}
\begin{tabular}{|l|c|c|c|c|c|} 
  \hline
  Paper & 
  \thead{Prob.\\Base Rate\\Adoption} & 
  \thead{Prob.\\Social\\Reinforcement} &
  \thead{Variable\\Time of\\Influence} &
  \thead{Threshold\\Model} &
  \thead{Variable Social\\ Reinforcement\\Threshold}\\ 
  \hline
  Current Paper &
  \textbf{Yes} &
  \textbf{Yes} &
  \textbf{Yes} &
  \textbf{Yes} &
  \textbf{Yes}\\ 
    \hline
  Centola \& Macy \cite{centola2007complex} &
  No &
  No* &
  No (SI only) &
  Yes &
  Yes\\ 
    \hline
  Sassine \& Rahmandad \cite{sassine2024does} &
  Yes &
  No &
  Yes &
  Yes &
  No\\ 
    \hline
  Eckles et al \cite{eckles2024long} &
  Yes &
  No* &
No* (SI only) &
  Yes &
  Yes\\ 
\hline
  De Kerchove et al. \cite{de2009role} &
  Yes &
  Yes &
  No (SIR only) &
  Yes &
  No\\ 
  \hline
  Zheng et al. \cite{zheng2013spreading} &
  Yes &
  Yes &
  No (SIR only)&
  No &
  NA\\ 
  \hline
O'Sullivan et al. \cite{o2015mathematical} &
  Yes &
  Yes &
  No (SIR only)&
  Yes &
  No\\ 
  \hline
  Keating et al.   \cite{keating2022multitype} &
  Yes &
  Yes &
  No (SIR only)&
  No & 
  NA\\ 
  \hline
  \end{tabular}
\caption{Comparison of modeling parameters to related literature. We only summarize key parameters as they pertain to our results so we exclude parameters that are not common with ours.
Asterisk (*) denotes that authors numerically simulate few instances with the given parameter, but not enough to capture its full variation.}
\label{tab:summary1}
\end{table}

\begin{table}
\renewcommand{\arraystretch}{1.2} 
\begin{tabular}
{|p{0.25\linewidth}|p{0.75\linewidth}|} 

\hline
  Paper & 
  Key Findings\\ 
  \hline
  Current Paper &
  \begin{compactitem}
        \itemindent=-17pt
        \item Four distinct regions of spread
        \item Tradeoff between reach and reinforcement
        \item Faster spread on clustered networks only a small region
        \item Degree and time of influence modulate impact of network 
    \end{compactitem}

 \\ 
    \hline
  Centola \& Macy 
 \cite{centola2007complex} &

\begin{compactitem}
        \itemindent=-17pt
        \item Socially reinforced behaviors spread better on clustered networks
\end{compactitem}
\\ 
    \hline
  Sassine \& Rahmandad 
 \cite{sassine2024does} &
    \begin{compactitem}
        \itemindent=-17pt
        \item Socially reinforced behaviors benefit from clustering when base rate adoption is low, degree is low, and time of influence is short
        \item Short times of influence stifle spread on random networks
    \end{compactitem}    

\\ 
\hline
  Eckles et al. 
 \cite{eckles2024long} &
    \begin{compactitem}
        \itemindent=-17pt
        \item Socially reinforced spread better on clustered when adoption is deterministic
        \item Even low probabilistic base rate flips result in favor of random ties (socially reinforced adoption rates remains deterministic) 
    \end{compactitem}
\\ 
\hline
  De Kerchove et al. \cite{de2009role} &
   \begin{compactitem}
        \itemindent=-17pt
        \item socially reinforced behaviors benefit from clustering when base rate adoption is low and social reinforcement is high
        \item as base rate rate adoption increases, clustering is beneficial with lower levels of social reinforcement
    \end{compactitem}
\\ 
  \hline
  Zheng et al.\cite{zheng2013spreading} &
   \begin{compactitem}
        \itemindent=-17pt
        \item Socially reinforced behaviors benefit from clustered networks when base rate adoption is low, social reinforcement it high, and network degree is low
        \item Identify four qualitatively distinct regions, but conclude results support findings of \cite{centola2007complex}.
    \end{compactitem}
\\ 
  \hline
O'Sullivan et al. \cite{o2015mathematical} &
   \begin{compactitem}
        \itemindent=-17pt
        \item The greater social reinforcement, the more beneficial clustering
        \item Authors conclude results support \cite{centola2007complex}
    \end{compactitem}
\\ 
  \hline
Keating et al. \cite{keating2022multitype} &
   \begin{compactitem}
        \itemindent=-17pt
        \item Socially reinforced behaviors benefit from clustering when base rate adoption is low and social reinforcement is high
        \item As base rate rate adoption increases, clustering is beneficial with lower levels of social reinforcement
    \end{compactitem}
\\ 
  \hline

\end{tabular}
\caption{Comparison to related literature. We only summarize key findings as they pertain to our results.}
\label{tab:summary2}
\end{table}

\clearpage
\singlespacing
\section{Comparing the Analytical Boundary of Spread to Numerical Results in Clustered Networks}\label{si_clust}

To determine regions of the $p_1 \leq p_2$ space where full spread is reached on random and clustered networks, we calculate the number of individuals, or ``initial adopters,'' who adopt from the seeds in the first $T$ time steps to estimate longer-term behavior.
Specifically, we derive the boundary in the $p_1 \leq p_2$ space where the number of initial adopters equals the number of seeds, above which full spread is likely reached.
For random networks, this boundary closely maps to the values of $p_1 \leq p_2$ where full spread will be reached.
For clustered networks, this boundary underestimates values of $p_1$ and $p_2$ where full spread is reached.

\begin{figure}
    \centering
    \includegraphics[width = \textwidth]{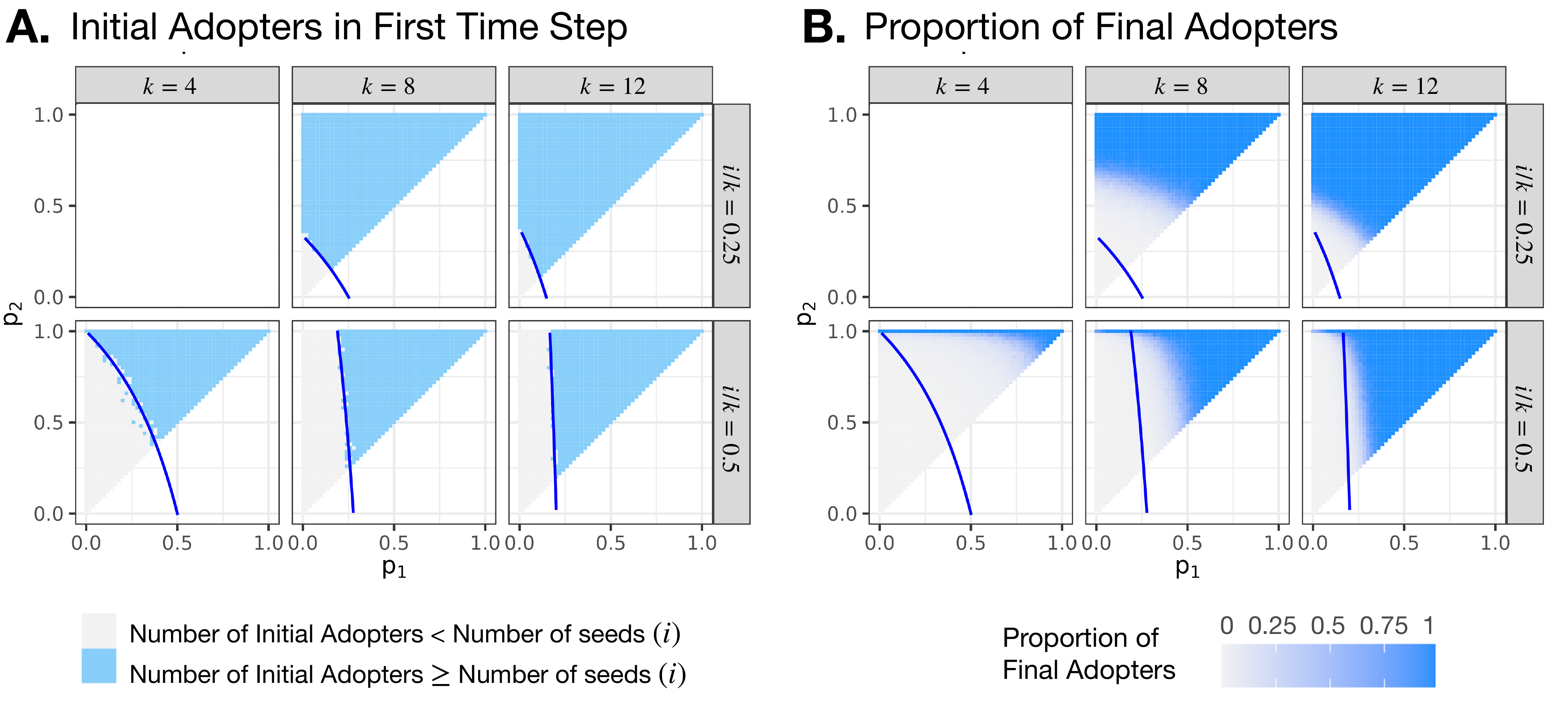}
    \caption{\textbf{Comparing initial and final spread on clustered networks to the theoretical boundary.} The analytical boundary (blue line) closely maps to the region in the $p_1 \leq p_2$ space where the number of initial adopters equals or exceeds the number of seeds, and underestimates the values of $p_1$ and $p_2$ where full spread is likely reached on clustered networks. This boundary also lowers for greater degree ($k$) and shifts right for greater social reinforcement thresholds ($i$), mirroring the effects of these parameters born out in numerical simulations.}
    \label{fig:si_cluster_net}
\end{figure}

\begin{figure}
    \centering
    \includegraphics[width = \textwidth]{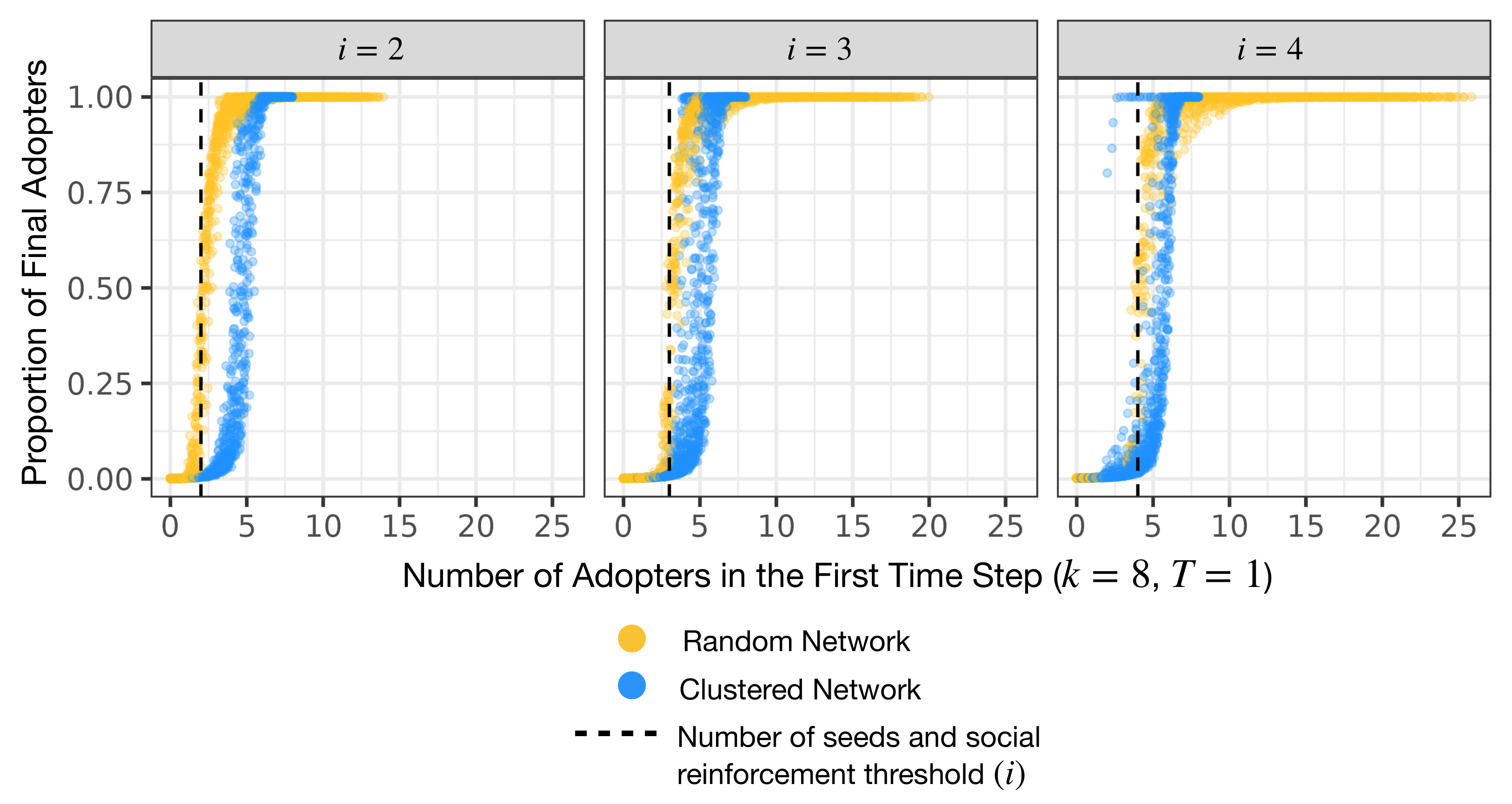}
    \caption{\textbf{Proportion of final adopters by the number of initial adopters in the first time step, for different social reinforcement thresholds (i).} For random networks, full spread is likely reached when the number of initial adopters is equal to or exceeds the number of seeds (indicated by black dashed lines). On clustered networks, most diffusion processes require even more initial adopters than the number of seeds for full spread to reached. In this figure, $k=8$ and $T = 1$ are used as example parameters.}
    \label{fig:si_i_plot}
\end{figure}

This discrepancy is due to the fact that while the analytical proof closely maps where in the $p_1 \leq p_2$ space the number of initial adopters equals the number of seeds, on clustered networks the region where the number of initial adopters equals or exceeds the number of seeds does not exactly match the regions where full spread is reached (Figure \ref{fig:si_cluster_net}).
Instead, full spread is reached for higher values of $p_1$ and $p_2$ than the area where the number of initial adopters equals or exceeds the number of seeds.
Even if there is a sufficient number of initial adopters, the diffusion process may still fail and not reach full spread (Figure \ref{fig:si_i_plot}).
Despite this imprecision, the analytical boundary allows us to accurately model how $p_1, p_2, k, T$ and $i$ affect spread on random and clustered networks.

\clearpage
\singlespacing
\section{Quantifying Regions of Spread with KS-tests}\label{si:ks}

\begin{figure}
    \centering
    \includegraphics[width = \textwidth]{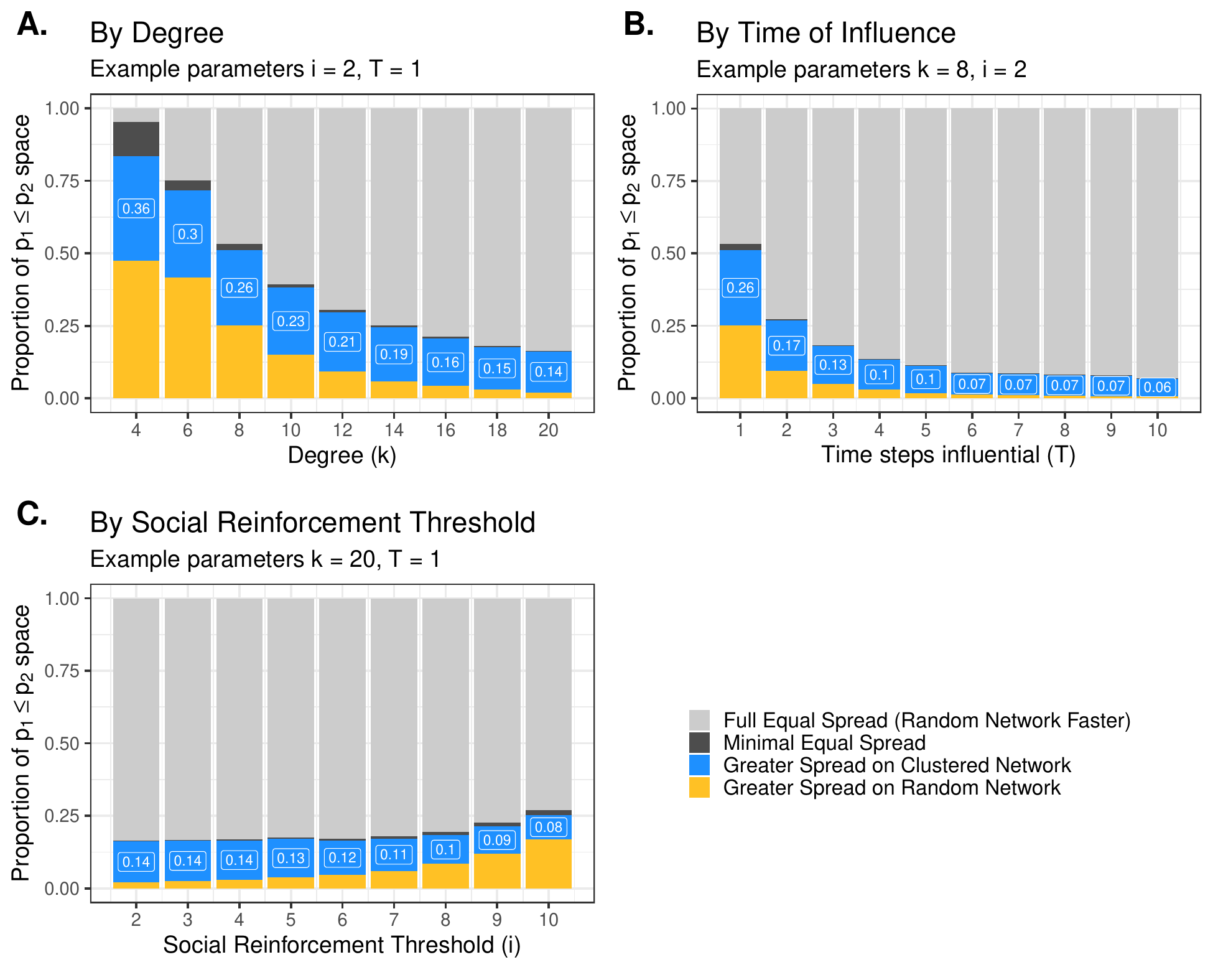}
    \caption{\textbf{Effect of Degree, Time of Influence, and Social Reinforcement Threshold
on Regions of Spread in the $p_1 \leq p_2$ Space Using KS Tests}. The proportion of the $p_1 \leq p_2$ where clustered networks outperform random networks (blue), random networks outperform clustered networks (orange), both network types have equal minimal spread (dark gray), or both network types have equal full spread (but faster on random networks, light gray) for varying degree (\textbf{A}), time of influence (\textbf{B.}), and social reinforcement threshold (\textbf{C.}).  One network type is considered to outspread the other if the proportion of adopters averaged over 100 simulations of the same parameter combination is greater for one network type and statistically significant at the $p<0.05$ for a two sided, two-sample, KS test.
Full equal spread is when $p\geq0.05$ and at least 60\% spread is reached for both network types. Minimal equal spread is when $p\geq0.05$ and less than 60\% spread is reached on either network type.
}
    \label{fig:si_ks}
\end{figure}

In the main text, we use a 5\% margin of difference in the average proportion of adopters to determine whether one network substantively outspreads the other. 
While this is a meaningful threshold, it is also arbitrary.
There may be regions where clustered networks consistently spread a behavior more than random networks with a margin of difference that is narrowly below the 5\% cutoff.
To account for this possibility, we reproduce Figure 3 (see Results) using a two-sided, two sample non-parametric Kolmogorov-Smirnov (KS) test instead of the 5\% difference in means (Figure \ref{fig:si_ks}).
Doing so provides a measure of statistical significance to compare random and clustered networks.
In this case, a clustered (random) network is considered to spread a behavior farther than a random (clustered) network if the proportion of final adopters averaged across the 100 trials on the clustered (random) network is greater than that of the random (clustered) network and the $p$-value from the KS test is less than a $0.05$ criterion.
If $p\geq0.05$, then the two network types are considered equal in how far they spread a behavior. 
If the proportion of final adopters is at least 60\% of the total number of individuals in the network on both random and clustered networks, the behavior is considered to attain full equal spread on both network types.
If the average proportion of final adopters is below 60\% on either network type, the behavior is considered to attain minimal equal spread on both network types.
For a majority of parameter combinations in the $p_1 \leq p_2$ space, the proportion of adopters, averaged over 100 trials, falls below 10\% or above 90\% (Figure \ref{fig:time_inf}).
This means the 60\% threshold is robust to other choices of a cutoff.

Using a KS test instead of a difference in means reveals similar results to that in the main results.
Regions where clustered networks outperform that of random networks constitute a minority of the total $p_1 \leq p_2$ space.
The region where clustered networks perform better than random networks decreases with greater degree, time of influence, and social reinforcement threshold.
As degree and time of influence increase, the area where clustered networks perform better contracts, while the area where full spread is reached on both networks equally (though at a faster rate on random networks) expands.
As the social reinforcement threshold increases, areas where clustered networks perform better decrease and areas where random networks perform better increase.

Overall, as the KS test captures differences in the average proportion of adopters within the 5\% margin, there tends to be proportionally more of the $p_1 \leq p_2$ space where one network outspreads the other compared to areas where there is either minimal equal spread or full equal spread.
The other main contrast between the difference in means and KS test methods is that clustered networks appear to do better for a much larger proportion of the space using the KS test method, especially when network degree is low.
For low degree cases in particular, when both $p_1$ and $p_2$ are small, there is a greater proportion of the $p_1 \leq p_2$ space where the average proportion of adopters on clustered networks is statistically greater than that of random networks (by overcoming the $p<0.05$ criterion), but this margin of difference is within 5\% of the total network size.

\clearpage
\singlespacing
\section{Minimal amount of social reinforcement needed for clustered networks to outperform random networks}

For each parameter combination, we calculate the minimum ratio of $p_2$ to $p_1$ ($p_2/p_1$) where a behavior spreads farther on a clustered network compared to a random network by a 0.1, 1, 5, and 10\% margin of difference of the total network size.
We do so by randomly sampling 10 simulated diffusion trials from both clustered and random networks for each parameter combination.
The proportion of adopters is averaged across the 10 trials, from which we calculate the difference in spread between clustered and random networks. 
We then select the smallest ratio of $p_2$ to $p_1$ among parameter combinations where behaviors spread faster by at least the specified margin of difference (0.1, 1, 5 and 10\% of the network size).
This process is repeated 1,000 times to produce a point estimate and 95\% confidence intervals.

For a 5\% difference in spread, the smallest ratio of $p_2$ to $p_1$ is when $p_2$ is nearly two times that of $p_1$ (minimum $p_2/p_1 =1.90$, 95\%CI = (1.87, 1.93) where $k=20$, $i=2$, $T=1$). This ratio decreases with smaller margins of difference, and increases with larger margins of difference. This ratio also increases with longer times of influence $T$ and greater social reinforcement threshold $i$.

\begin{figure}
    \centering
    \includegraphics[width = \textwidth]{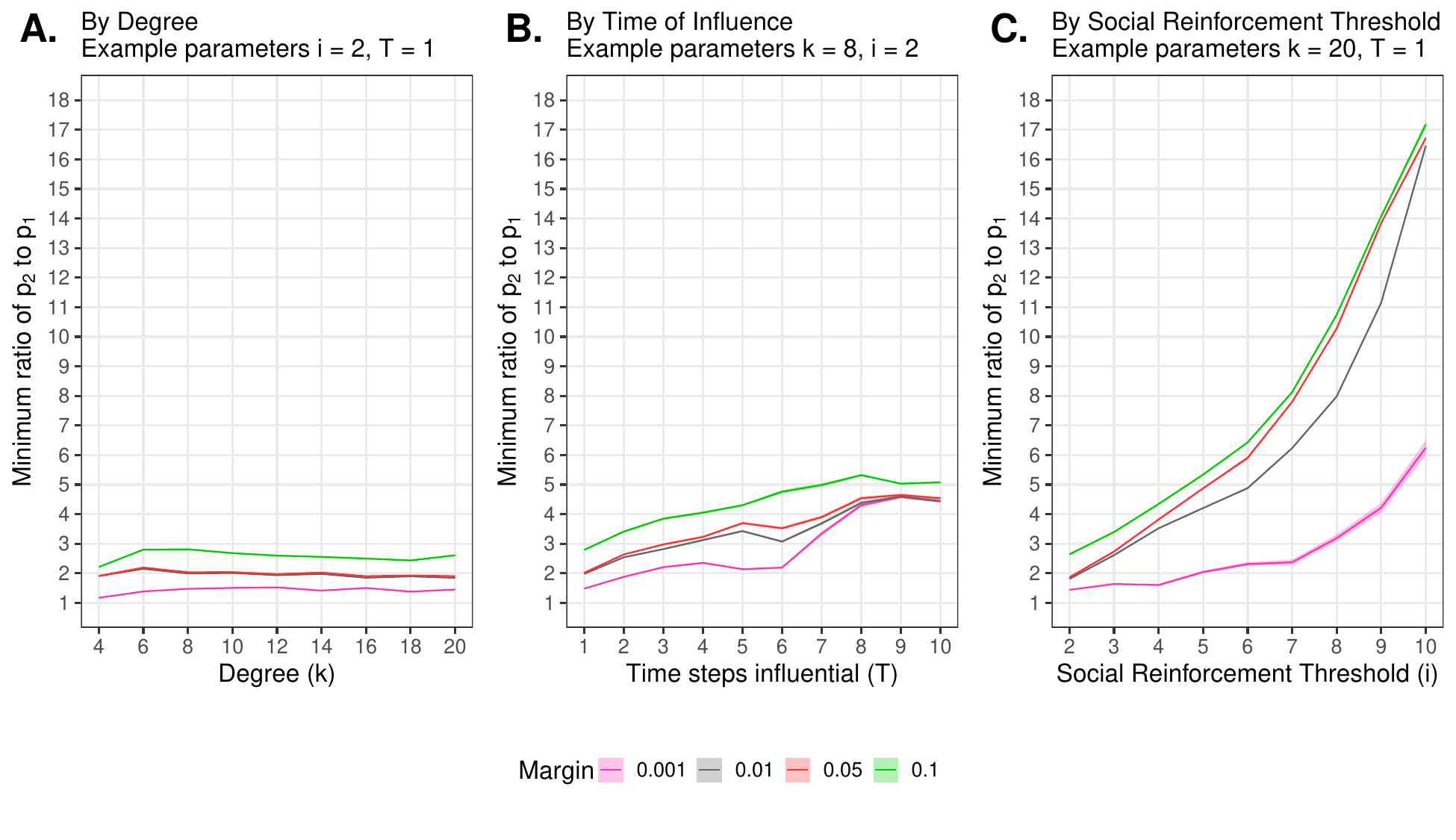}
    \caption{\textbf{Minimum $p_2/p_1$ ratio where clustered networks outperform random networks.} Statistic is calculated for differing degree ($k$, \textbf{A.}), time of influence ($T$, \textbf{B.}) and social reinforcement threshold ($i$, \textbf{C.}) for 0.1, 1, 5, and 10\% differences in spread. Ribbons show bootstrapped 95\% confidence intervals based on 1,000 replications comparing 10 randomly sampled values from the random and clustered network each.}
    \label{fig:si_p2_p1}
\end{figure}

\clearpage
\section{Speed of Spread}\label{si_speed}

Here, we show that random networks diffuse a behavior faster than random networks in the region where behaviors reach full spread on both random and clustered networks (Figure \ref{fig:speed}).
Speed of spread is measured as the number of time steps required for a diffusing behavior to be adopted by 60\% of all individuals in the network, averaged over 100 independent trials per parameter combination.
For a majority of parameter combinations in the $p_1 \leq p_2$ space, the proportion of adopters, averaged over 100 trials, falls below 10\% or above 90\% (Figure \ref{fig:time_inf}).
This means that which parts of the space that are considered to reach full equal spread is robust to a large variation in the specific level of network saturation we choose.
We can expect that measuring time to achieve a greater level of network saturation (such as 90\%) will result in longer spreading times, while the reverse is true for lower levels of saturation.
Regardless of the exact saturation level, random networks spread behavior faster than clustered networks in this region.

\begin{figure}
    \centering
    \includegraphics[width = \textwidth]{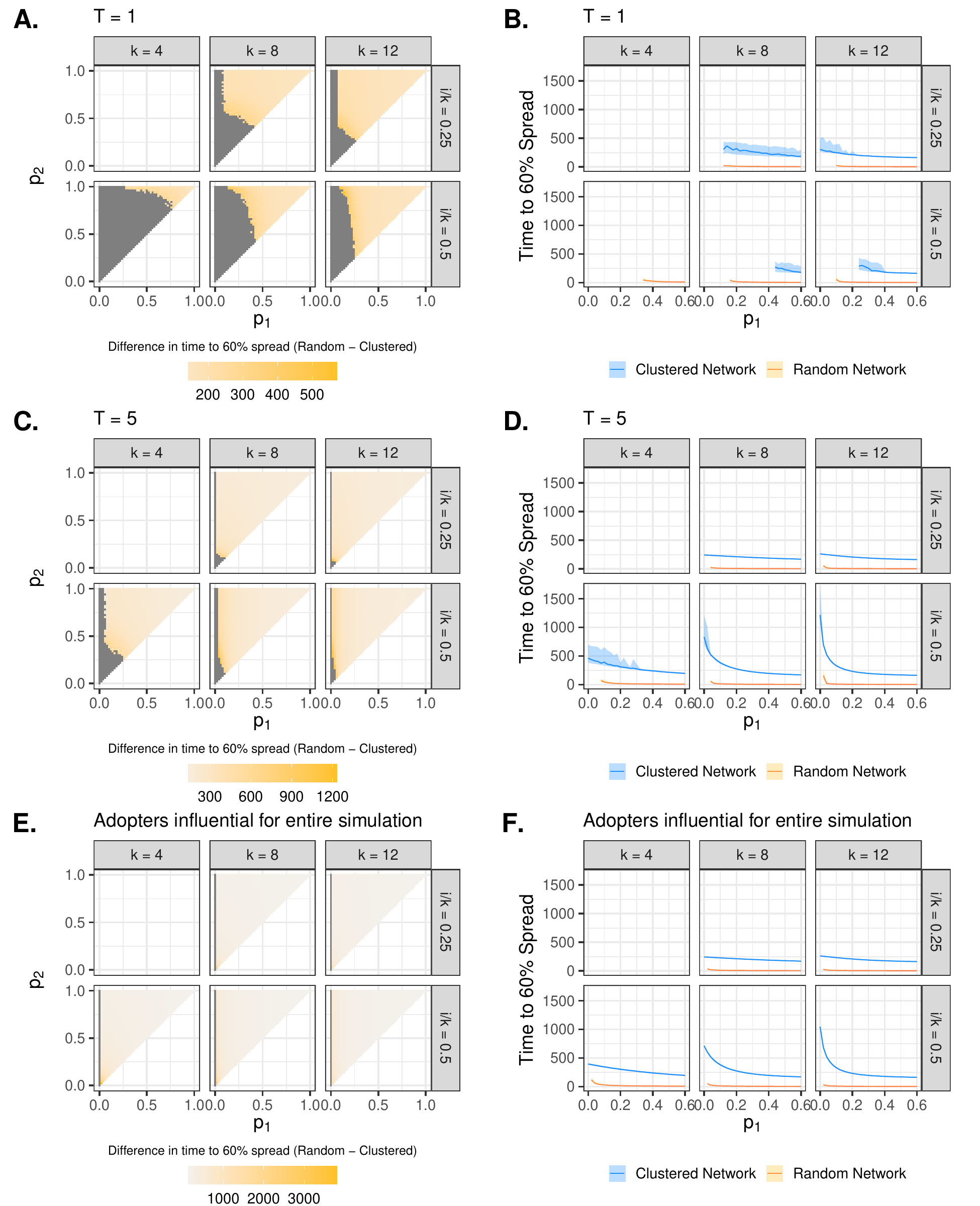}
\end{figure}
\cleardoublepage
\begin{center}
    \captionof{figure}{\textbf{Speed of spread on random and clustered networks.} Time to 60\% spread (60\% of the network adopts the behavior) for the entire $p_1 \leq p_2$ space when $T = 1$ (\textbf{A.}), $T = 5$ (\textbf{C.}), and when adopters remain influential for entire simulation  (\textbf{E.}). 
    Time to 60\% spread when $p_2 = 0.6$ for the entire $p_1 \leq p_2$ space when $T = 1$ (\textbf{B.}), $T = 5$ (\textbf{D.}), and when adopters remain influential for entire simulation (\textbf{F.}).  Ribbons show 95\% confidence intervals.
    Parameter combinations where the proportion of adopters, averaged over 100 trials, is less than 60\% on either network type are excluded (dark gray in the figures of the full $p_1 \leq p_2$ space).  
    Random networks are always faster than clustered networks in regions of the $p_1 \leq p_2$ space where 60\% spread is reached on random networks.}
    \label{fig:speed}
\end{center}

\clearpage
\section{Role of Time of Influence}\label{si_time_inf}

\begin{figure}
    \centering
    \includegraphics[width = \textwidth]{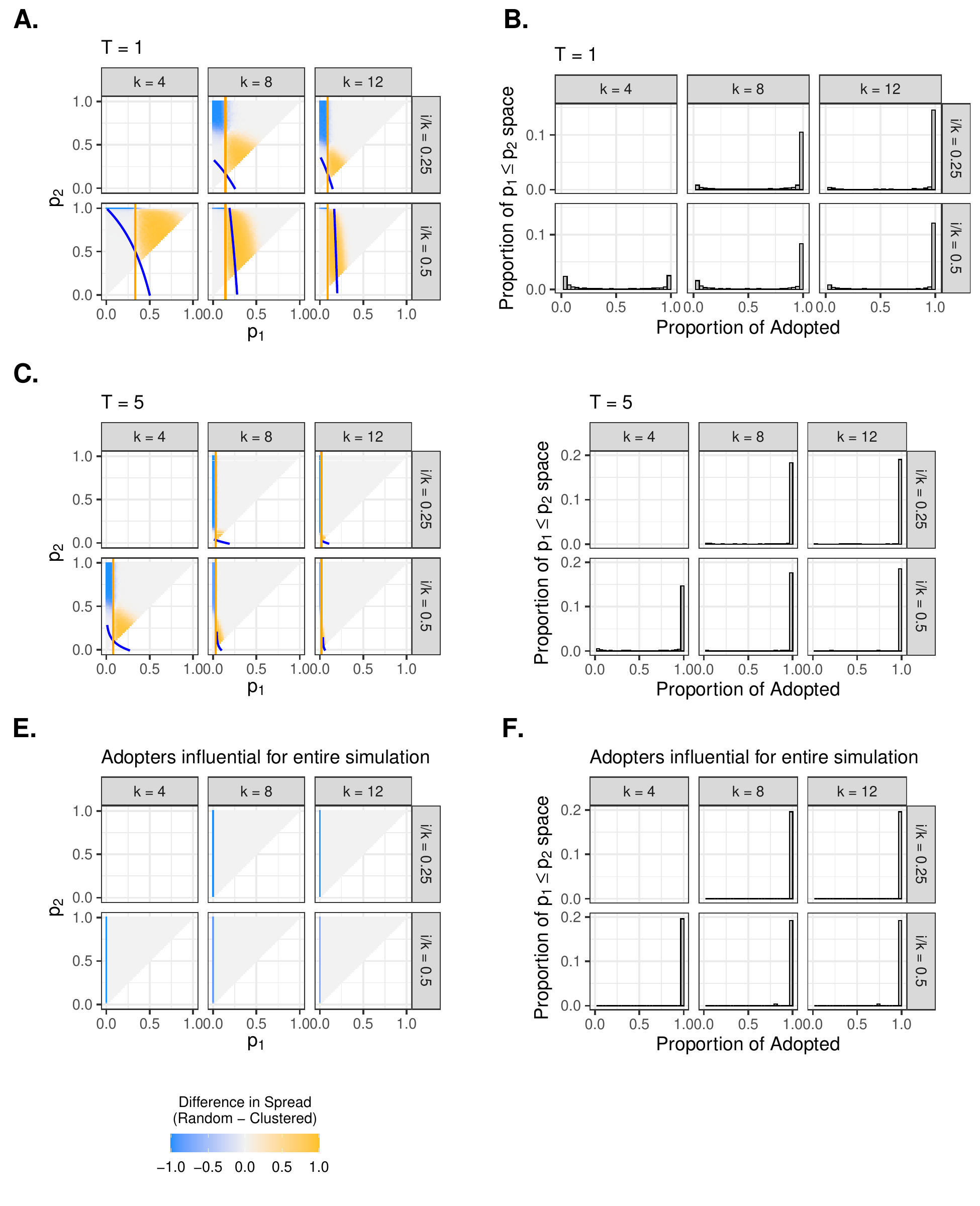}
    \caption{\textbf{Effect of Time of Influence on Spread}. 
    Regions of spread in numerical results for $T = 1$(\textbf{A.}), $T = 5$ (\textbf{C.}), and when adopters remain influential for entire simulation  (\textbf{E.})
    Distribution of proportion of adopters, averaged over 100 trials for $T = 1$(\textbf{B.}), $T = 5$ (\textbf{D.}), and when adopters remain influential for entire simulation  (\textbf{F.}).
    }
    \label{fig:time_inf}
\end{figure}

We show the effect of a longer time of influence on the values of $p_1$ and $p_2$ for which behavior spreads on random and cluster networks, for differing degree and social reinforcement thresholds (Figure \ref{fig:time_inf}).
Even increasing the time of influence from one time step to five time steps dramatically decreases the space where either network outperforms the other.
As time of influence increases, the probability that an influential individual will successfully influence their neighbor also increases, facilitating spread on both random and clustered networks.
This means greater time of influence increases the area where full spread can be reached on both clustered and random networks, increasing the region where behavior reaches full spread on both network types (though at a faster rater on random networks) and decreasing the area where either network outperforms the other.

\clearpage
\section{Two-Dimensional Clustered Networks}\label{si_dim}

The ring lattices used in the main analysis represent a narrow case of clustered networks. Here, we evaluate how the $p_1 \leq  p_2$ space changes when we use a two-dimensional ``Moore'' lattice instead of a one-dimensional ring lattice. Specifically, we compare ring and Moore lattices with example parameters $k =8$, $i = \{2,4\}$, and $T=1$.

\begin{figure}
    \centering
    \includegraphics[width = \textwidth]{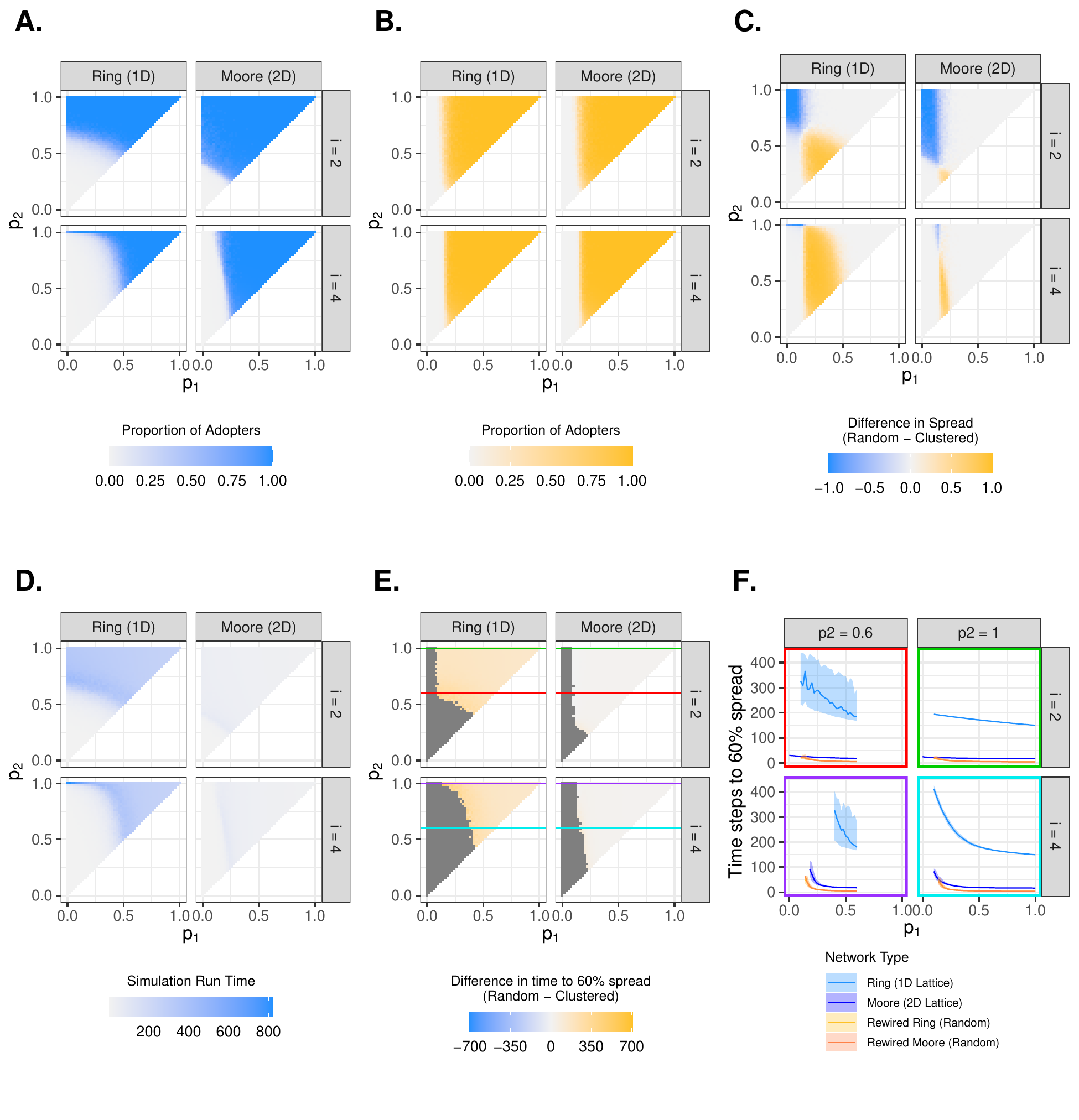}
    \caption{\textbf{Comparing Ring (1-Dimensional) and Moore (2-Dimensional) Networks, with example parameters $k = 8, T = 1$}. Average proportion of adopters on clustered lattice networks(\textbf{A.}) and randomly rewired lattices (random networks, \textbf{B.}). \textbf{C.} Difference in spread between clustered and rewired random networks. \textbf{D.} Simulation run time (Moore lattice faster). \textbf{E.} Difference in time to 60\% spread between random and clustered networks. Parameter combinations where the proportion of adopters, averaged over 100 trials, is less than 60\% on either network type are excluded (dark gray). \textbf{F.} Time to 60\% spread when $p_2 = 0.6$ (red and cyan lines/boxes) and $p_2 = 1$ (green and purple lines/boxes) for ring, Moore, rewired ring, and rewired Moore networks. Ribbons show 95\% confidence intervals. Points excluded where the proportion of adopters is less than 60\% or $p_1>p_2$. }
    \label{fig:dim_compare}
\end{figure}

In ring lattices, the continued propagation of a behavior rests heavily on being able to pass through a bottleneck of individuals buffering those individuals who have already adopted and those who have yet to be exposed. 
On higher dimensional networks, this bottleneck is less pronounced as there are more directions (rather than just along the ring in the one-dimensional case) for a behavior to spread along.
Should some individuals fail to adopt, stifling diffusion in one direction, there are more directions for the diffusion process to continue by circumventing ``dormant'' individuals.
As a result, Moore lattices diffuse a behavior with lower values of $p_1$ and $p_2$ compared to ring lattices, increasing the area within the $p_1 \leq p_2$ space where behaviors spread on clustered networks (Figure \ref{fig:dim_compare}A).
There is no difference between spread on random networks rewired from ring lattices compared to random networks rewired from Moore lattices (Figure \ref{fig:dim_compare}B).
In comparing spread on random and clustered networks, Moore lattices increase Regions 1 and 3 where clustered networks are advantageous, and where spread is reached on both random and clustered networks respectively, at the expense of Region 2 where random networks are advantageous (Figure \ref{fig:dim_compare}C).

Moore networks also spread behavior faster than ring lattices (Figure \ref{fig:dim_compare}D), narrowing the difference in speed between random and clustered networks in the region where full spread is reached on both network types (Figure \ref{fig:dim_compare}E \& F).
Despite this, random networks still spread behavior faster.

\clearpage
\section{Effect of Different Seeding Strategies}\label{si_seed}

In the main article, we seed both random and clustered networks by randomly selecting one individual and $i-1$ neighbors as the initial seeds. This ``neighbor seeding'' is done to ensure a consistent seeding strategy across network types and that there is at least some likelihood that an initial adopter (individual adopting from the seeds) can adopt at the above threshold, socially reinforced rate. This is especially important for the spread of behaviors when the base rate, $p_1 =0$. 
However, there are other seeding strategies that could potentially lead to different results. Here, we evaluate two other seeding strategies. 

\begin{figure}
    \centering
    \includegraphics[width = \textwidth]{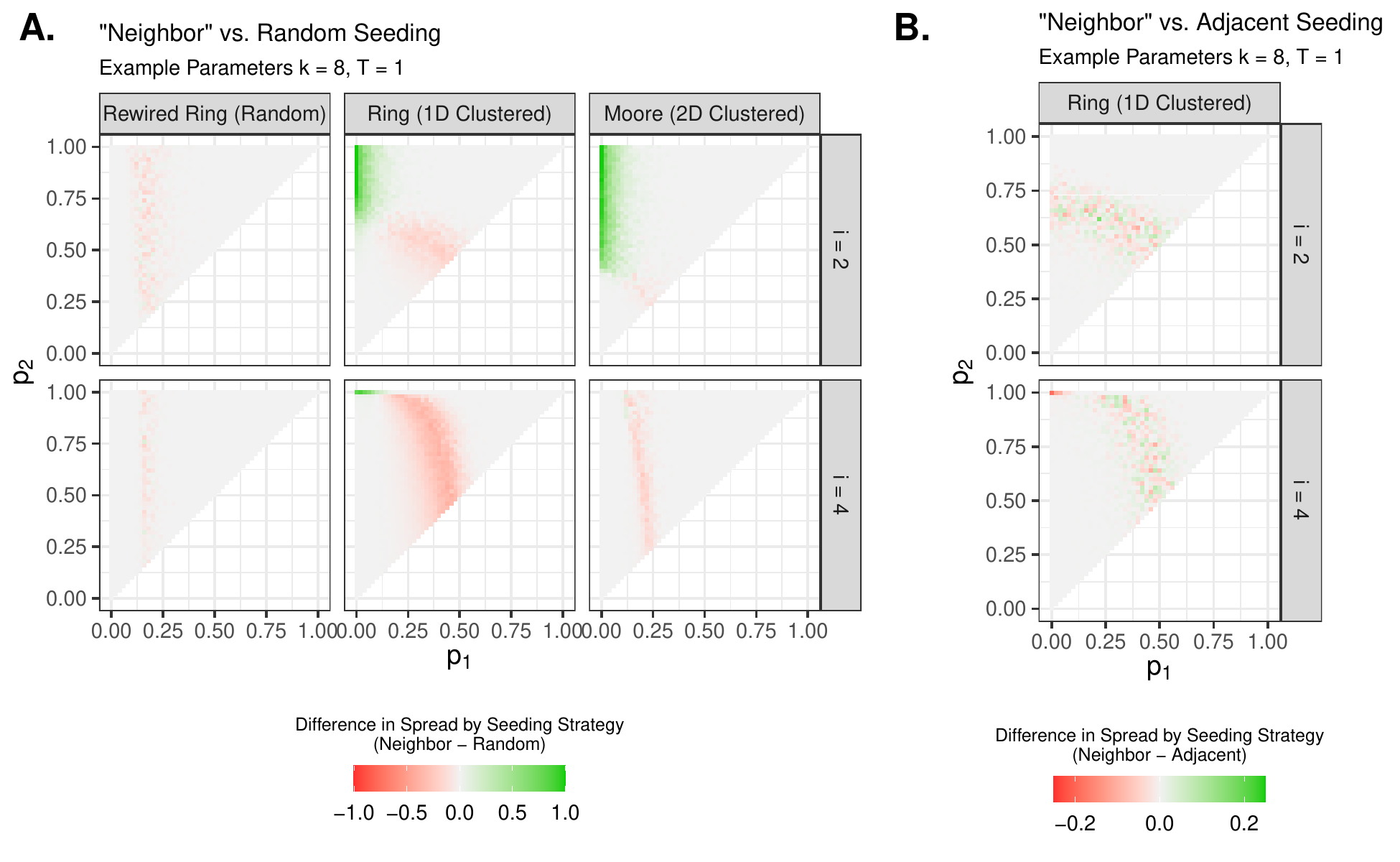}
    \caption{\textbf{Effect of Alternative Seeding Strategies, with example parameters $k = 8, T = 1$}. \textbf{A.} Difference in proportion of adopters, averaged over 100 trials, between ``neighbor'' seeding (select one individual and $i-1$ neighbors) and random seeding (select $i$ random individuals) on rewired ring (random), ring, and Moore networks. Green areas are where there is more spread using ``neighbor'' seeding. Red areas are where there is more spread using random seeding.
    \textbf{B.} Difference in the proportion of adopters between ``neighbor'' seeding (select one individual and $i-1$ neighbors) and ``adjacent'' seeding (select $i$ individuals and $i-1$ directly adjacent neighbors) on ring lattices. Green areas are where there is more spread using ``neighbor`` seeding.
    Red areas are where there is more spread using ``adjacent'' seeding.}
    \label{fig:seeding}
\end{figure}

First, we consider random seeding, where $i$ random individuals are selected as the seeds (Figure \ref{fig:seeding}A). Compared to neighbor seeding, random seeding stifles spread on clustered networks in more deterministic areas of the space, when $p_1$ is low. This is because it is less likely that any initial adopter will be in contact with multiple seeds, precluding adoption at the socially reinforced rate. Thus, $p_1$ must be high enough that a behavior can be transmitted from the seeds to the initial adopters on non-redundant ties alone.
In contrast, random seeding facilitates spread on clustered networks for greater $p_1$, since the base rate is high enough to make use of non-redundant ties.
On random networks, there is a slight advantage to random seeding. This is because there is a larger set of individuals in contact with the seed nodes with random seeding compared to neighbor seeding. With more independent chances (one for every individual in contact with a seed), $p_1$ can be slightly lower for diffusion to still occur.

Finally, as a special case for the ring lattice, networks can be seeded by choosing nodes that are directly adjacent to each other (Figure \ref{fig:seeding}B). We find this makes very little difference for overall spread.

\clearpage
\section{Effect of Heterogeneous Adoption}\label{si_het}

In the main paper, we model adoption homogeneously. For a given $p_1$ and $p_2$, every individual has the same base rate and socially reinforced adoption probability. 
How do our results hold up when individual adoption is modeled as heterogeneous instead of homogeneous?

To test this, we make one change to the original model. 
Each time an individual is exposed to an influential neighbor, rather than adopting at $p_1$ or $p_2$, the individual adopts at a rate drawn from a normal distribution centered at either $p_1$ or $p_2$.
With this adjustment, adoption is heterogeneous within individuals.
By tuning the standard deviation of this normal distribution, $\sigma$, we can control the degree of heterogeneity. 
When $\sigma = 0$, adoption is homogeneous and collapses onto our main analyses; as $\sigma$ increases, adoptions becomes increasingly more heterogeneous.

\begin{figure}
    \centering
    \includegraphics[width = \textwidth]{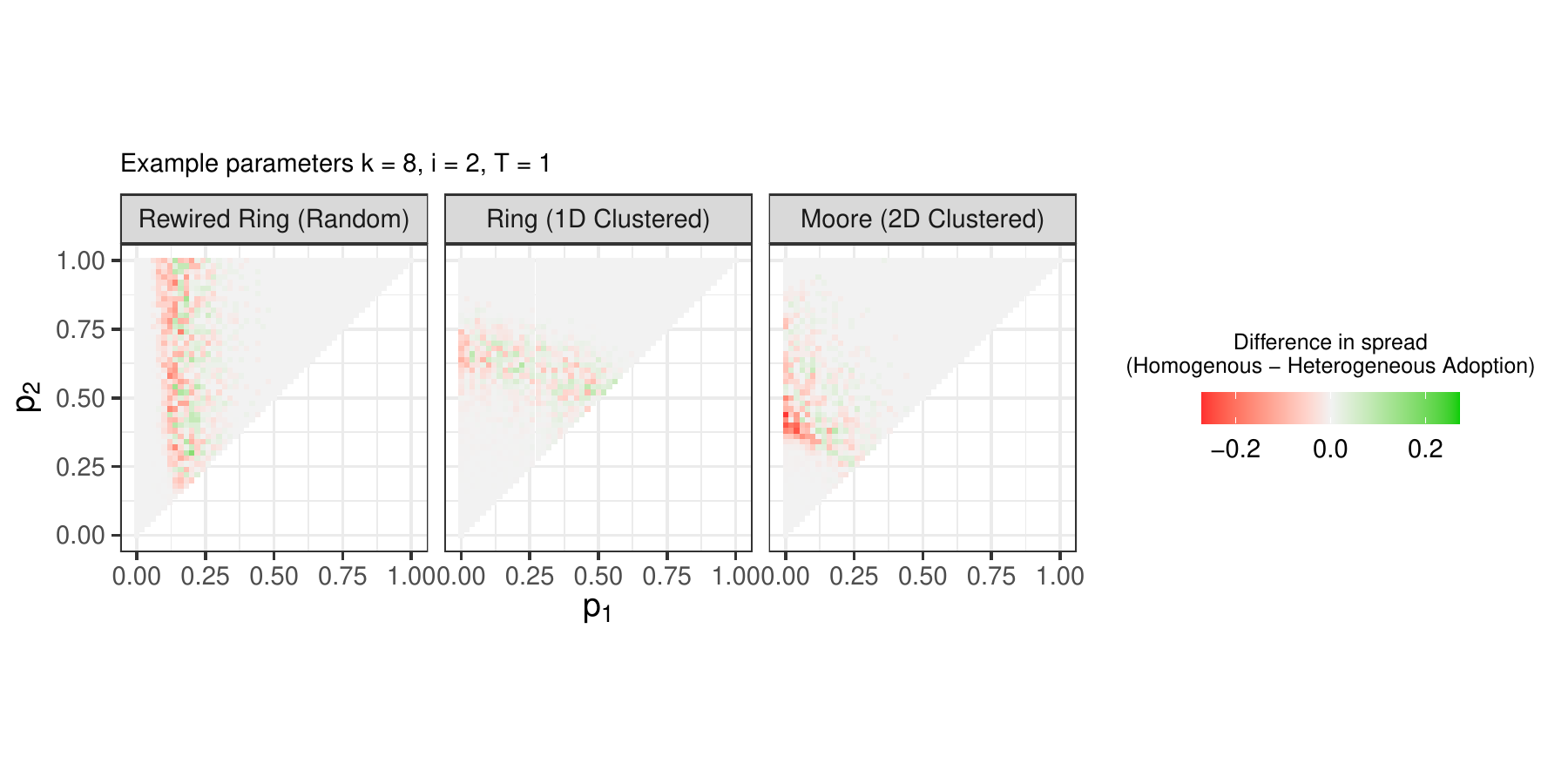}
    \caption{\textbf{Heterogeneous adoption}. Difference in the proportion of adopters, averaged over 100 trials, between homogeneous adoption with $\sigma = 0$ and  heterogeneous adoption with $\sigma = 0.1$ on rewired ring (random), ring, and Moore networks. Green areas are where there is more spread with homogeneous adoption. Red areas are where there is more spread with heterogeneous adoption. We use example parameters $k=8, i = 2, T =1$}
    \label{fig:het}
\end{figure}

We find that incorporating heterogeneous adoption has little effect on the results (Figure \ref{fig:het}). 
On random networks, heterogeneous adoption slightly increases spread for lower $p_1$ values.
Among clustered networks, we observe a similar effect where spread increases among lower $p_1$ and $p_2$ values. 
For all cases, heterogeneous adoptions serve to slightly expand the borders where either clustered or random networks spread. 
This is because heterogeneous adoptions allow for a certain amount of adoption above the specified $p_1$ and $p_2$ values.
That being said, even for fairly large amounts of heterogeneity, where $\sigma = 0.1$ (such that for a given adoption rate $p$, 95\% of adoptions fall between a 0.4 point range on a 0 to 1 scale), we observe at most a 27\% ($p_1 = 0, p_2 = 0.44$ on the Moore lattice) difference in spread for a very small area of the space.

\clearpage
\section{Effects of Intermediate Rewiring}\label{si_rewiring}

Real world networks are rarely uniformly clustered or uniformly random. Instead, they are often characterized by locally clustered neighborhoods connected by random ties \cite{watts1998collective}. To account for some of the heterogeneity in local clustering, we repeat our analyses with different amounts of degree preserving rewiring (See Methods) from 0 of the edges rewired (equivalent to the clustered network in the main analysis) to all of the edges rewired (equivalent to the random network in the main analysis). 

First, we consider the effects of intermediate rewiring on how far a behavior diffuses through the network (Figure \ref{fig:rewiring}A \& B).
Especially when time of influence is short as opposed to long (for instance $T = 1$ as in Figure \ref{fig:rewiring}A), recall that there is a large amount of heterogeneity in whether a behavior reaches full spread on either clustered on random networks (See Results). 
There are regions where behaviors spread better on clustered networks (Region 1), spread better on random networks (Region 2), spread on both networks types (Region 3), or spread minimally on both network types (Region 4).
The effect of rewiring will largely depend on which region a behavior falls within for a given parameter combination.
We focus on several illustrative cases of points that fall in each of these regions, but we expect these findings to generalize to other points within the $p_1 \leq p_2$ space.
We also limit our analyses to $k= 8$,$n = 2000$ ,$i = 2$, but expect similar results with different parameterizations.

For behaviors that fall definitively within Regions 1 or 2 where one network type clearly outperforms the other, the effect of rewiring is monotonic (barring a certain amount of noise).
For behaviors in Region 1 ($p_1 = 0, 0.001,0.01, 0.1$ and $p_2 = 1$ on both the ring and Moore lattices in Figure \ref{fig:rewiring}A), the more edges that are rewired, the less a behavior spreads.
With low $p_1$ and high $p_2$, behaviors in this region can only spread through highly clustered, redundant ties. 
Rewiring disrupts this local redundancy, hindering spread.
For smaller $p_1$ values, even small amounts of rewiring can disrupt the spreading process.
As $p_1$ increases, behaviors are more likely to spread through rewired random ties, so the negative effect of rewiring on spread is less pronounced.
For instance, when $p_1 = 0, 0.001, \text{ or }0.01$, a behavior can no longer spread when half of the edges are rewired.
However, for larger $p_1$, such as $p_1 = 0.1, 0.2$, while rewiring dampens spread, these behaviors can still spread on fully rewired random networks.
The opposite case is true for behaviors in Region 2 where random networks outperform clustered networks (for instance, $p_1 = 0.2, p_2 = 0.25$ on both ring and Moore lattices in Figure \ref{fig:rewiring}A).
More rewiring quickly increases spread.

More complex dynamics are observed for points that fall around the boundaries of these regions. 
Notably, the points $p_1 = 0.1, p_2 = 0.5$ and $p_1 = 0.2, p_2 = 0.5$ are close to the boundaries of Regions 1, 2, and 4 when diffusing on ring lattices.
Here, we observe a boost in spread at an intermediary level of rewiring that is greater than both the non-rewired and fully rewired networks.

When adopters remain influential for the entire simulation, the $p_1 \leq p_2$ space reduces to clustered networks better spreading a behavior (Region 1) when $p_1 = 0$ and a behavior spreading on both random and clustered networks, but a faster rate on random networks everywhere else (Region 3; Figure \ref{fig:rewiring}F). 
This means that aside from $p_1 = 0$ where rewiring quickly dampens spread, for all other values full spread can be attained regardless of the level of rewiring (Figure \ref{fig:rewiring}B).

Next, we turn to the effect of rewiring on the speed of spread, which we measure as the time for 90\% of the network to adopt.\footnote{Simulations that don't reach 90\% spread are excluded. We expect similar results with lower level of network saturation, with the exception that the difference in time will be less pronounced.} 
We first consider the case where adopters remain influential for the entire simulation.
Across both ring and Moore lattices, the dominant pattern we observe is that rewiring speeds up spread (Figure \ref{fig:rewiring}D).
The only exceptions to this occur when $p_1$ is near deterministic, and among the example points we consider, when $p_1 < 0.1$.

On ring lattices, when $p_1 = 0$, the time to 90\% spread decreases with additional rewiring, but only because spread is entirely stifled by the time 30\% of edges have been rewired (Figure \ref{fig:rewiring}B). 
In this case, rewiring is not speeding up the time to spread, but preventing spread altogether.
When $p_1$ is slightly increased to 0.001, intermediate levels of rewiring up to 10\% of edges speed up diffusion.
Once more than 10\% of edges are rewired, additional rewiring slows down spread. 
This is similar to the ``small world window'' observed in \cite{centola2007complex}.
The small world window is less pronounced when $p_1 = 0.01$, and disappears altogether when $p_1 = 0.1$.
For $p_1 = 0.1, 0.2$, any amount of rewiring speeds up diffusion.
Most notably, even rewiring just 0.001\% of edges (2 edges in the 2000 node network used here) dramatically speeds up diffusion.

Different values of $p_2$ affect spreading speed when there is little to no rewiring, but have no effect on speed when the network is fully rewired.
This is consistent with our main results showing that spread on random networks does not depend on $p_2$ since  individuals are rarely adopting at the socially reinforced rate.
As $p_2$ increases, spread on networks with minimal rewiring speeds up.
For higher $p_2$ (for example $p_2 = 1, 0.5$), this boost in spread is great enough that non-rewired clustered networks are faster than random networks for small $p_1$ where the small world window exists.
For smaller $p_2$ (like $p_2 = 0.25$), the boost in speed is not great enough; rewired random networks still spread a behavior faster even with the small world window.

Moore lattices behave similarly to ring lattices with one key difference. As higher dimensional lattices diffuse behaviors faster than one-dimensional lattices (see SI Section \ref{si_dim}), the baseline speed for networks with minimal rewiring is much faster. 
As there is no effect of Moore as opposed to ring lattices on the speed of fully rewired networks (ring and Moore lattices rewire to the same regular random networks), using Moore lattices generally tilts the favor towards clustered networks.
Since the baseline speed on clustered networks is faster, the small world window disadvantages random networks more than in the case of the ring lattice.
We also observe a small world window when $p_1 = 0$, which was not previously the case with the ring lattice.
Despite this, once $p_1$ is high enough that the small world window disappears (for $p_1$ values as small $0.1$ in our example points), we observe a monotonic effect of rewiring that favors random networks.
More rewiring speeds up diffusion rather than slowing it down.

When time of influence is short, many diffusion processes never saturate to at least 90\% of the network (for instance when $p_2 = 0.25$). For the points in the $p_1 \leq p_2$ space that do, we observe similar behaviors to when adopters always remain influential. 
There is a small word window effect that is more pronounced on Moore lattices which disappears as $p_1$ increases.
For $p_1 \geq 0.1$, rewiring monotonically speeds up diffusion (Figure \ref{fig:rewiring}C).

\begin{figure}
    \centering
    \includegraphics[width = \textwidth]{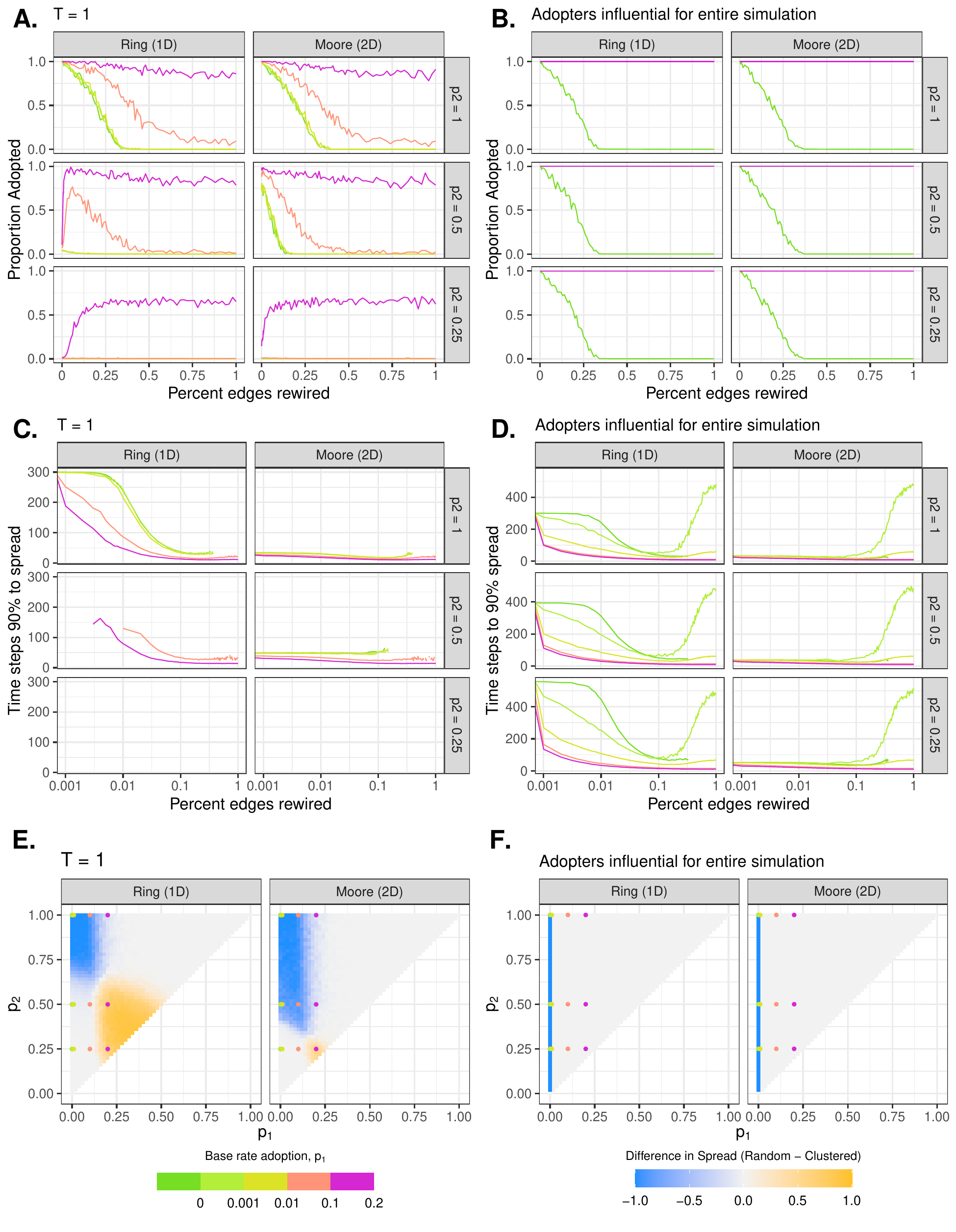}
\end{figure}
\cleardoublepage

\begin{center}
  \captionof{figure}{
    \textbf{Effect of Intermediate Rewiring with example parameters $k=8, i =2$}. The effect of rewiring on the average proportion of final adopters when $T=1$(\textbf{A.}) and when adopters remain influential for the entire simulation (\textbf{B.}), time steps to 90\% spread when $T=1$(\textbf{C.}) and when adopters remain influential for the entire simulation (\textbf{D.}), for example points $p_1 = \{0, 0.001, 0.01, 0..1, 0.2\}, p_2 = \{0.25, 0.5, 1\}$.
    Example points overlaid on the relevant $p_1 \leq p_2$ space showing regions of spread from numerical results for $T = 1$ (\textbf{E.}) and when adopters remain influential for the entire simulation $T=1$(\textbf{F.}). Points with less than 90\% spread are excluded from the panels showing speed (making $p_2 = 0.25$ blank in panel C).}
    \label{fig:rewiring}
\end{center}

\clearpage
\section{Centola and Macy 2007 Replication}\label{si_cm07}

\begin{figure}
    \centering
    \includegraphics[width = \textwidth]{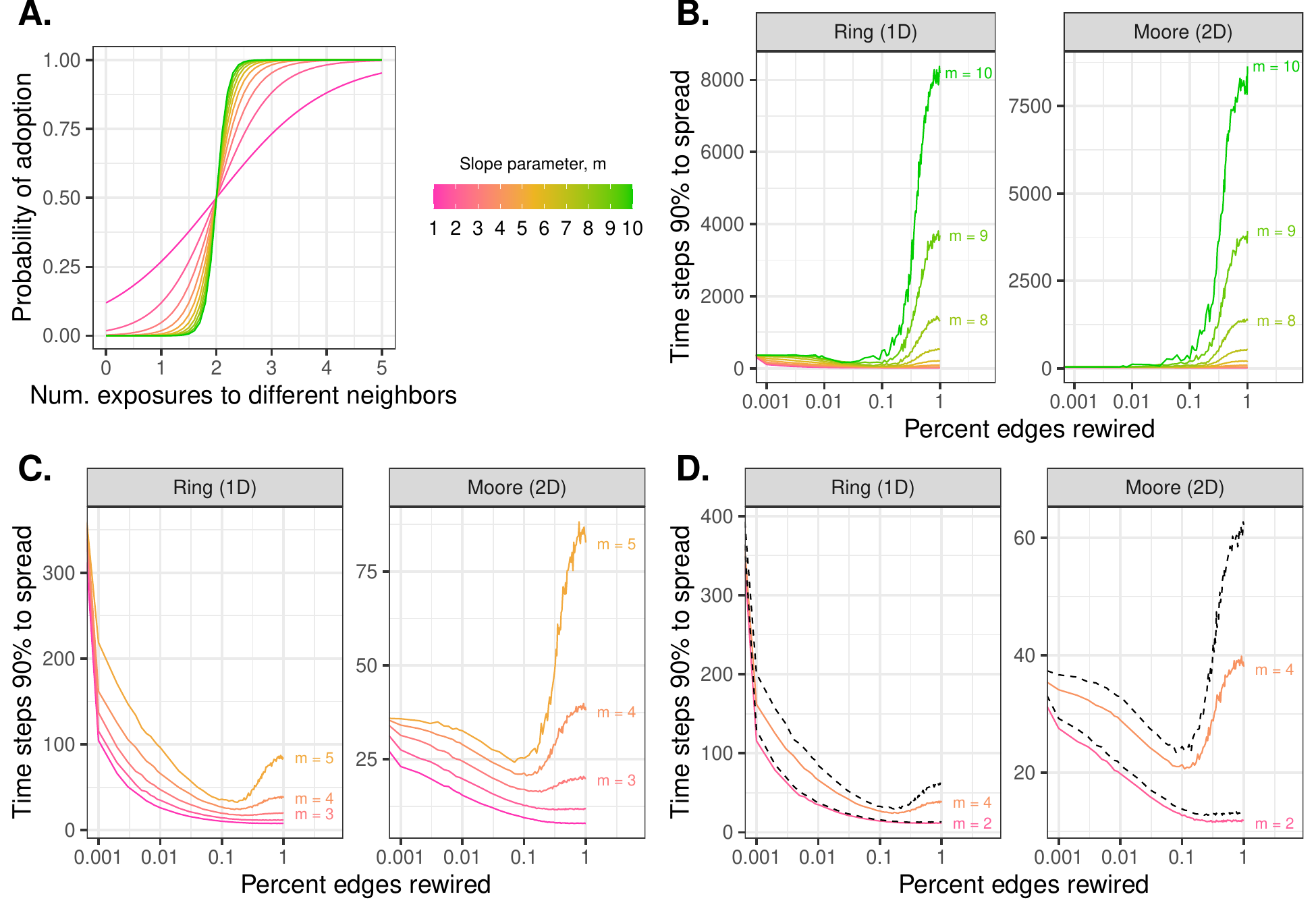}
    \caption{\textbf{Effect of Rewiring on Reach and Spread using a Cumulative Logistic Function}. \textbf{A.} Cumulative logistic function used to model the probability of adoption with different values of the slope parameter $m$.\textbf{B.} Time to 90\% spread from $m=1$ to $m=10$ for different levels of rewiring. \textbf{C.} Time to 90\% spread for from $m=1$ to $m=5$ for different levels of rewiring (for better visibility). \textbf{D.} Comparison between the effects of rewiring using a cumulative logistic function (colored lines) as opposed to a step function (dashed lines). The dashed line closely matching $m=4$ shows the effects of rewiring for $p_1 = 0.01, p_2 = 0.5$. The dashed line closely matching $m=2$ shows the effects of rewiring for $p_1 = 0.1, p_2 = 0.5$. We use parameters $k=8, i = 2$, and adopters remain influential for the entire simulation.}
    \label{fig:mc07}
\end{figure}

To more closely compare our results with previous work, we conduct a replication of Centola and Macy \cite{centola2007complex}. Specifically we replicate the ``stochastic threshold'' robustness check, which examines one case of a behavior with probabilistic adoption, but find that it spreads faster on clustered networks.
In a key departure from our model, instead of using a step function, Centola and Macy model the adoption trajectory as a cumulative logistic function (``sigmoid", based on that in \cite{macy1990learning}; Figure \ref{fig:mc07}A). 
The adoption probability, $p(c)$, is expressed as a function of the number of different influential neighbors an individual has been exposed to since the start of the simulation $c$, such that
\[p(c) = \frac{1}{1+e^{(2-c)m}}\].
When individuals are exposed to two different influential neighbors, $p(2) = 0.5$
The probability of adoption approaches 0 when individuals have exposure to less than two neighbors, and approaches 1 when individuals have exposure to more than two neighbors.
The ``slope'' parameter $m$ tunes how deterministic $p(c)$ is.
When $m$ is high (such as $m = 10$ in Figure \ref{fig:mc07}A), $p(1)$ is very close to 0 and $p(3)$ is close to 1. 
As $m$ decreases, $p(1)$ increases from 0 and $p(3)$ decreases from 1. 
Regardless of $m$, $p(2) = 0.5$.
The primary difference between modeling adoption as a sigmoid function as Centola and Macy do or as a 2-parameter ($p_1$, $p_2$) step function as we do is that $p(3) \to 1$ in the sigmoid function, and $p(3) = p_2$ in the step function we use.

We explore the diffusion of behaviors parameterized by the sigmoid function for values of $m$ from 1 (least deterministic) to 10 (most deterministic).
Consistent with the modeling in Centola and Macy \cite{centola2007complex}, once individuals adopt, they remain influential for the entire simulation, which is equivalent to having a long or ``infinite" time of influence.
Like the original paper, we evaluate the speed of spread as edges are rewired from the original clustered (either ring or Moore lattice) resulting in fully rewired regular random networks.
We use the same rewiring scheme as the rest of the paper \cite{gkantsidis2003markov}.

Despite using a slightly different function to model adoption probability, we observe similar patterns to the effects of rewiring on the step function (Figure \ref{fig:mc07}B \& C).
On ring lattices, for large values of $m$ where $p(1)$ is close to 0, the effect of rewiring is similar to when $p_1$ is small ($p_1 = 0,0.001, 0.1 $  in Figure \ref{fig:rewiring}D).
Small amounts of rewiring offer some boost in speed, but additional rewiring slows spread creating the ``small world window.''
The more deterministic the behavior (greater $m$), the more rewiring slows diffusion down, disadvantaging random networks.
However, as a behavior becomes less deterministic (for instance, $m = 2$ ), additional rewiring speeds up spread rather than slowing it down.
A similar effect is observed with Moore lattices, with the exception that the baseline speed for clustered networks with no rewiring are faster than that of ring lattices.
As a result, for some $m$ parameters where we previously observed a small world window, the effect of rewiring is instead a monotonic increase in time to saturate the network, slowing down spread.

Finally, as a direct comparison, we compare behaviors parameterized by the sigmoid function that most closely match the example points used in the rewiring analysis in SI Section\ref{si_rewiring} (Figure \ref{fig:mc07}D). 
When $m = 4$, $p(1) = 0.018$, which is close to $p_1 = 0.01$.
When $m = 2$, $p(1) = 0.12$, which is close to $p_1 = 0.1$.
Plotting $m=4$ with $p_1 = 0.01, p_2 = 0.5$, and $m=2$ with $p_1 = 0.1, p_2 = 0.5$ show qualitatively similar results.
Behaviors with more deterministic base rate adoption ($p_1$ in our step function, $p(1)$ in the cumulative logistic function) spread faster on clustered networks and are slowed down by rewiring.
Behaviors with less deterministic base rate adoption spread faster with additional rewiring, making random networks more advantageous.

\clearpage
\section{Centola 2010 Replication}\label{si_c2010}

A prominent empirical test of how clustering affects the spread of socially reinforced behaviors is the field experiment by Centola \cite{centola2010spread} (ref [21] in the main body). In this experiment, Centola studied the adoption of a health behavior (joining an online health forum) among participants in artificial online communities that were either structured as clustered or random networks. In both cases, participants connected to a randomly chosen seed node were sent an email to join an online health forum. If individuals joined the forum, each of their neighbors in the network would also receive an email inviting them to join, continuing the diffusion process. Centola found that diffusion was faster and farther among those embedded in clustered networks.

The modeling choices in our main analysis diverge somewhat from the empirical setting of Centola \cite{centola2010spread} potentially making it more difficult to generalize our results. To make our results directly comparable, we conduct a synthetic replication of Centola's experiment as closely as possible, based on available information. Centola's study uses smaller networks, two-dimensional rather than one-dimensional lattices, and only seeds networks with one randomly chosen individual rather than a cluster of adjacent neighbors. 
Accordingly we replicate our main analysis with clustered networks that are either hexagonal lattices with 128 individuals ($k = 6, n = 128$) or Moore lattices with 144 individuals ($k = 8, n = 144$), corresponding to the $B-F$ trials in Centola's study. Both network types are wrapped on tori so there are no boundary conditions. 

Identical to our main paper, random networks are created by rewiring the lattice networks through a series of connected double edge swaps. For each parameter combination in the $p_1 \leq p_2$ space, we simulate a diffusion process 100 times, where we create a new rewired random network for each trial. We choose a social reinforcement threshold of 2 to best match the Centola's empirical results where the likelihood of joining the health forum was highest when individuals were exposed to two adopting neighbors. We select a short time of influence ($T = 1$ and $T =2$) to reflect the fact that adopters influenced their neighbors through a single email notification, which presumably can quickly become buried under other notifications. We also test the effect of individual heterogeneity. That is, for a given $p_1, p_2$ combination, every time an individual is exposed to an influential neighbor they adopt at a rate drawn from a normal distribution with either $p_1$ or $p_2$ as the mean. Adjusting the standard deviation ($\sigma$) of the distribution can modulate how heterogeneous individuals are, with $\sigma =0$ meaning that individuals are homogeneous.

\begin{figure}
    \centering
    \includegraphics[width = \textwidth]{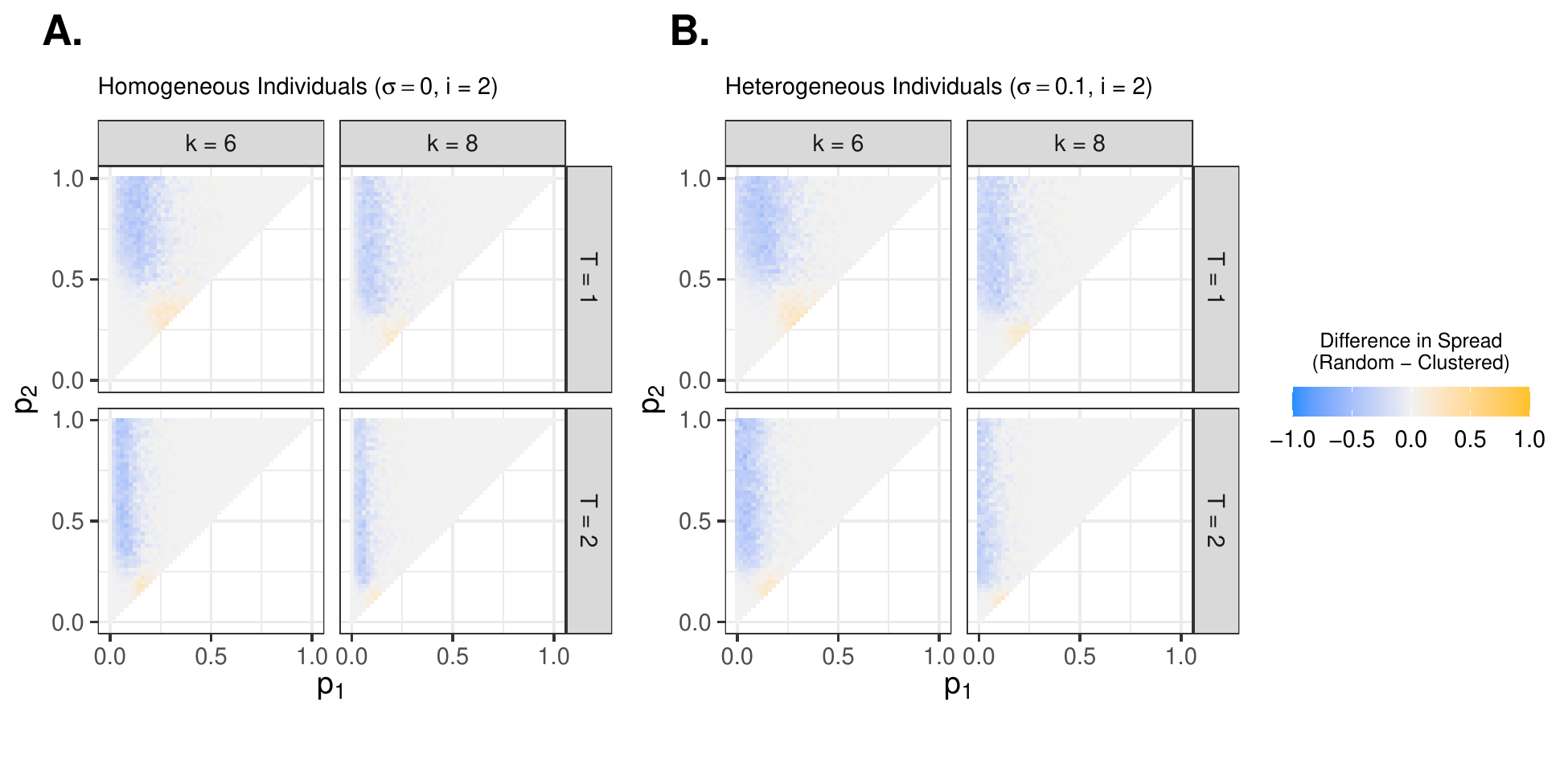}
    \caption{\textbf{Regions of spread from numerical simulation using exact networks from Centola 2010}. Difference in the proportion of adopters, averaged over 100 trials, between random and clustered networks where adoption is modeled homogeneously($\sigma = 0$ \textbf{A.}) and heterogeneously ($\sigma = 0$ \textbf{B.}). Random networks spread farther in orange regions and clustered networks spread farther in blue regions. The networks spread a behavior equally in the light gray regions. We use $i = 2$ to best match empirical results of the paper. }
    \label{fig:c2010_main_compare}
\end{figure}

With these modifications, we see that the effect of clustering on behaviors in the $p_1 \leq p_2$ space remains largely consistent with our main results (Figure \ref{fig:c2010_main_compare}). We recover the four distinctive regions where clustered networks are advantageous when the base rate adoption is relatively low and social reinforcement is relatively high. 
Since network degree is low and time of influence is short, clustered networks are beneficial in larger areas of the space, but see that even increasing network degree from 6 to 8 or time of influence from 1 to 2 time steps narrows the area where clustered networks are exclusively beneficial, consistent with out findings.
By virtue of using two-dimensional lattices, clustered networks are exclusively beneficial for lower levels of social reinforcement compared to one-dimensional ring-lattices, increasing the region where clustering is advantageous. 
Moreover, two-dimensional lattices spread a behavior faster than one-dimensional lattices (SI Section \ref{si_dim}).
Seeding the networks with just one randomly selected node also decreases the likelihood of spread on clustered network in increasingly deterministic regions of the space when the base rate of adoption $p_1$ is close to 0.
In fact, by seeding a network with just a single seed, it would be impossible for diffusion to happen without some positive amount of below threshold adoption.

Heterogeneity only has an effect for the narrow band of behaviors at the borders of each region where either clustered networks or random networks are exclusively beneficial.
This means that behaviors can spread on random networks for slightly lower levels of $p_1$ and on clustered networks, behaviors can spread when $p_1 = 0$, which was impossible in the homogeneous case when the diffusion process was only seeded with one individual. 
This is most apparent when contrasting the full adoption curves for homogeneous (Figure \ref{fig:si_c2010_ts_0sd}) and heterogeneous (Figure \ref{fig:si_c2010_ts_0.1sd}) individuals when $p_1 = 0$.
See SI Section \ref{si_dim}, \ref{si_seed}, and \ref{si_het} for evaluations of the individual effects of two-dimensional lattices, different seeding strategies, or heterogeneity respectively.

Centola's investigation only involved the comparison of six clustered networks to six random networks. As our simulated results are averaged over 100 trials, there is the likelihood that Centola's results represent an edge case in the space of behaviors we sample. Specifically, there are three areas of uncertainty that may impact the results: The particular node that is chosen as the seed, the particular random networks that are used, and the inherent stochasticity of the diffusion process when adoption is probabilistic
As we do not have access to any of these details, we can look at the variation in our results to assess the possibility that Centola may have arrived at differing conclusions given a specific parameter combination of base rate and socially reinforced adoption probabilities.

Beginning with variation in diffusion processes, Figure \ref{fig:si_c2010_ts_0sd} and \ref{fig:si_c2010_ts_0.1sd} show full adoption trajectories for all 100 trials for differing values of $p_1$ and $p_2$. As both $p_1$ and $p_2$ increase, the variation in trajectories narrows significantly. For high values of $p_1$ and $p_2$, all simulations reach full spread, and at a clearly faster rate on random networks. It is only among intermediary levels of $p_1$ and $p_2$ where larger variation is observed. For these parameters, there are clearly cases where a particular trial on a clustered (random) network could be observed to outperform a particular trial on a random (clustered) network even though one network type outperforming the other is observed as the dominant pattern (for instance, when $p_1 = 0.2$, and $p_2$ is between 0.4 and 0.8).

Supposing Centola's empirical results fall within an intermediary range of $p_1$ and $p_2$ subject to large variations across diffusion processes, could the choice of random network or seed have an effect?
To answer this, we simulate 100 diffusion trials on the same network with a different seed for each trial for the parameter combination $p_1 = 0.2, p_2 = 0.6$. We do this over ten different randomly rewired networks (Figure \ref{fig:si_c2010_rand_seed}B). 
We repeat this process using the same rewired network and the same seed for ten different networks and seeds (Figure \ref{fig:si_c2010_rand_seed}C).
See that across the ten conditions choosing the same network but different seed, as well as the ten conditions choosing the same network and same seed, there is no observable difference to that of the original where different networks and seeds are used for all 100 trials (Figure \ref{fig:si_c2010_rand_seed}A).
Since we choose a parameter combination that maximizes variation in diffusion trajectories, this suggests that the choice of a particular seed or network has little effect on the results. 

Altogether, our synthetic replication shows that the results of Centola reflect our main findings. There are regions in the $p_1 \leq p_2$ space where socially reinforced behaviors with probabilistic adoption spread only on clustered networks. However, for many different levels of $p_1$ and $p_2$ there are also results showing spread exclusively on random networks, spread on both networks types, or no spread at all. Centola's choice of two-dimensional networks, networks with low degree, and the transmission of a behavior with a presumably low time of influence all serve to maximize the region where clustered networks are beneficial. Replicating this study with a behavior with a longer time of influence, or on networks with a greater degree would require behaviors to be increasingly deterministic (low $p_1$, high $p_2$) in order for clustered networks to remain exclusively beneficial. 

\begin{figure}
    \centering
    \includegraphics[width = 9in, angle=90]{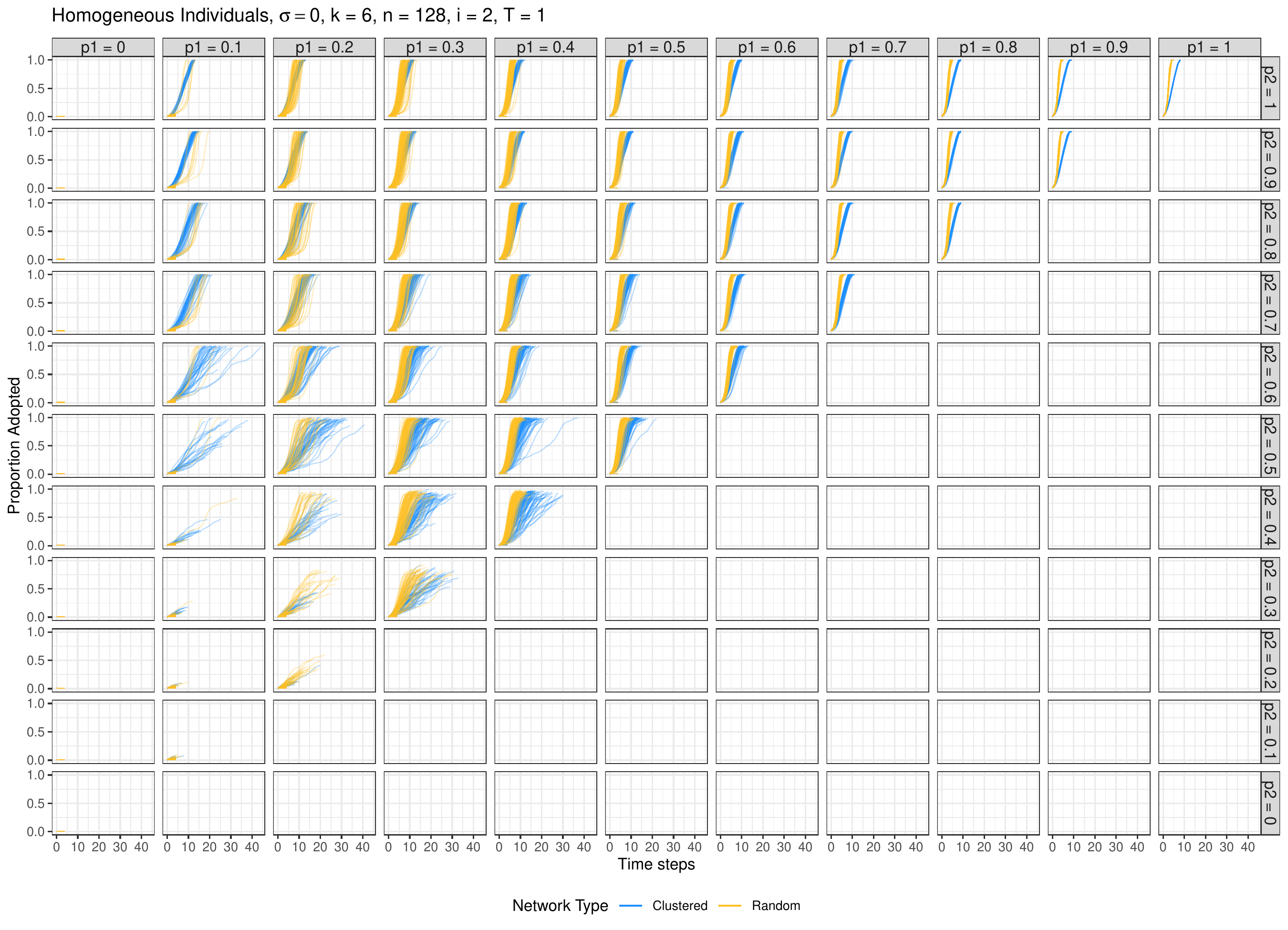}
\end{figure}
\cleardoublepage
\begin{center}
    \captionof{figure}{\textbf{Full time series of adoption on random and clustered networks with homogeneous adoption}. Each line represents one independent diffusion process on random (orange) and clustered (blue) networks. There are 100 trials (lines) for each $p_1, p_2$ parameter combination. Example parameters $\sigma = 0,k = 6,n = 128, i = 2, T =1$ for a subset of $p_1$ and $p_2$ values. }
    \label{fig:si_c2010_ts_0sd}
\end{center}

\begin{figure}
    \centering
    \includegraphics[width = 9in, angle=90]{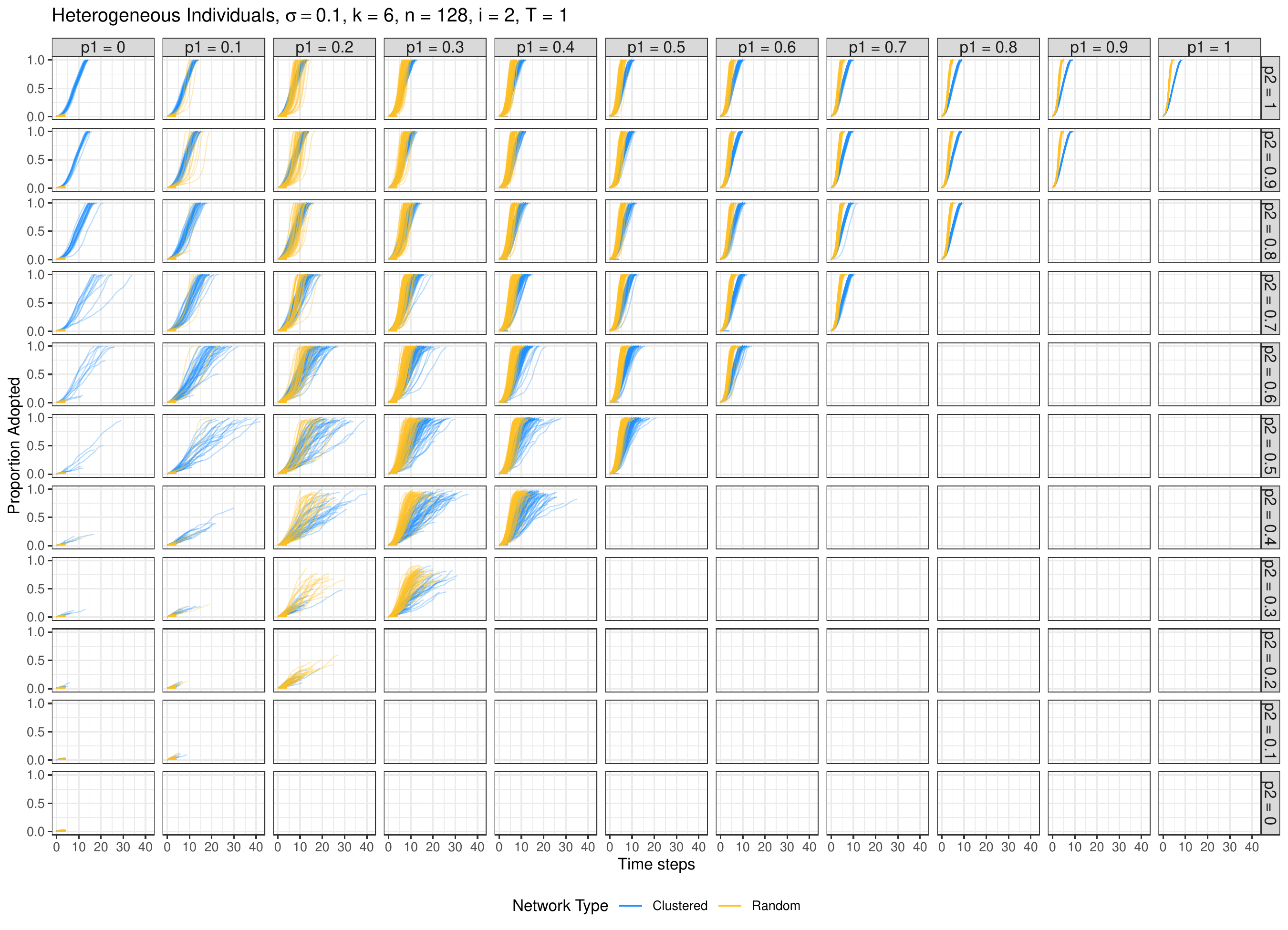}
\end{figure}
\cleardoublepage
\begin{center}
    \captionof{figure}{\textbf{Full time series of adoption on random and clustered networks with heterogeneous adoption}. 
     Each line represents one independent diffusion process on random (orange) and clustered (blue) networks. There are 100 trials (lines) for each $p_1, p_2$ parameter combination. Example parameters $\sigma = 0.1,k = 6,n = 128, i = 2, T =1$ for a subset of $p_1$ and $p_2$ values. }
    \label{fig:si_c2010_ts_0.1sd}
\end{center}

\begin{figure}
    \centering
    \includegraphics[width = \textwidth]{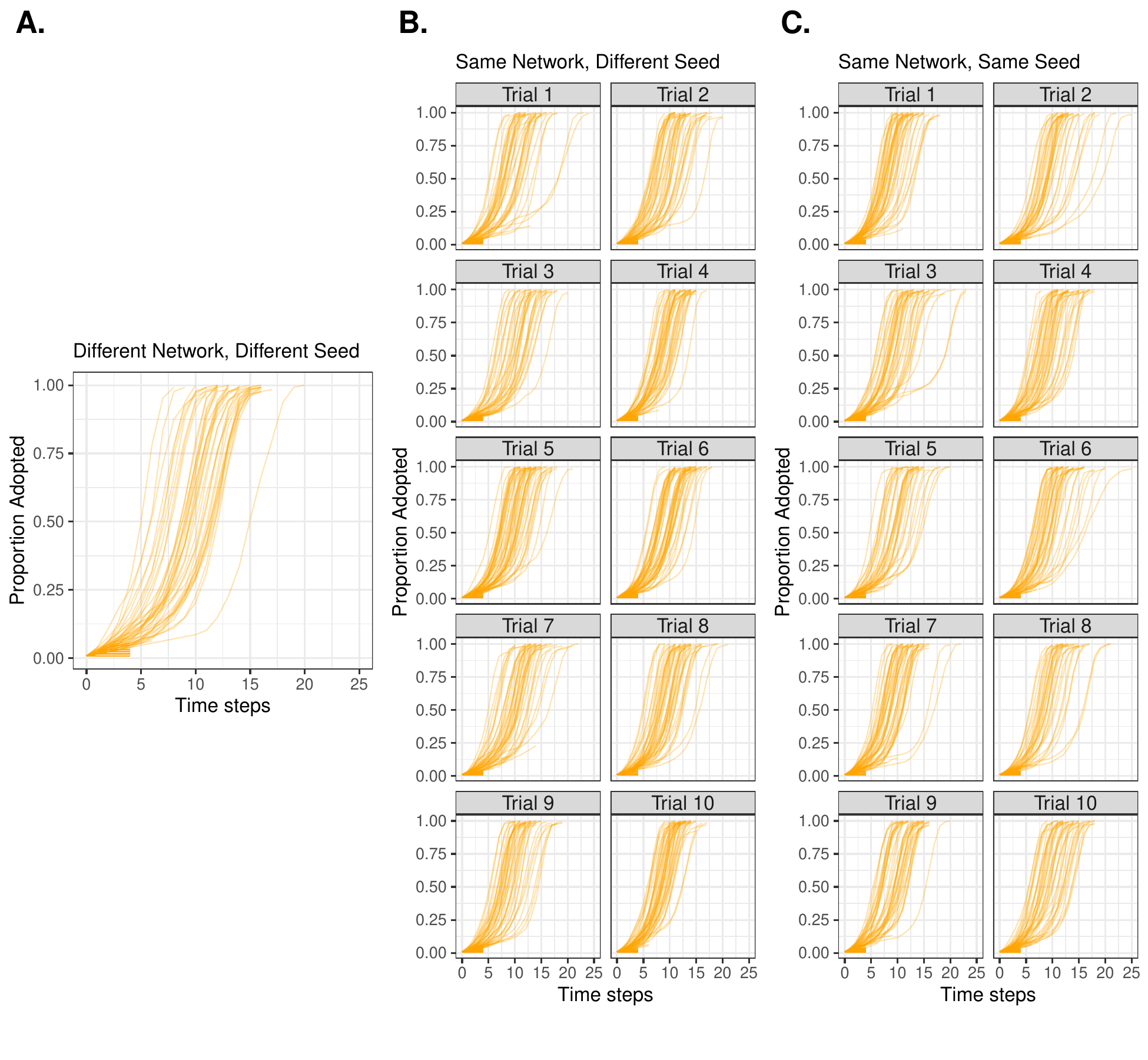}
    \caption{\textbf{Full time series of adoption on random networks for example parameters $p_1 = 0.2, p_2 =0.6,k = 6, ,n = 128, i = 2, T =1, \sigma = 0.1$}. \textbf{A.} 100 diffusion simulations of spread on random networks with a different random network and different seed for each simulation. \textbf{B.} 100 diffusion simulations of spread on random networks with the same random network across all 100 simulation but different seed for each simulation. We repeated this 10 times with a different random network.
    \textbf{B.} 100 diffusion simulations of spread on random networks with the same random network and same seed across all 100 simulation. We repeated this 10 times with a different random network and different seed.}
    \label{fig:si_c2010_rand_seed}
\end{figure}

\clearpage
\singlespacing
\bibliographystyle{ieeetr}
\bibliography{references_all.bib} 

\end{document}